\documentclass[letterpaper,journal]{IEEEtran}

\usepackage{amsmath,amsfonts,amssymb}
\usepackage{graphicx}
\usepackage{booktabs}
\usepackage{pifont}
\usepackage[table]{xcolor}
\usepackage{pgfplots}
\usepackage{cite}
\usepackage{url}
\usepackage{xspace}

\def\etal{\textit{et al.}\xspace}

\newcommand{\cmark}{\ding{51}}%
\newcommand{\xmark}{\ding{55}}%

\pgfplotsset{compat=1.18}
\usepgfplotslibrary{groupplots}
\definecolor{iorange}{RGB}{255,149,0}
\definecolor{ired}{RGB}{255,59,48}
\definecolor{igreen}{RGB}{52,199,89}
\definecolor{icyan}{RGB}{50,173,230}
\definecolor{iblue}{RGB}{0,122,255}
\definecolor{igray}{RGB}{142,142,147}
\definecolor{imimi}{RGB}{200, 180, 255} 

\title{Reconstruct! Don't Encode: Self-Supervised Representation Reconstruction Loss for High-Intelligibility and Low-Latency Streaming Neural Audio Codec}

\author{Junhyeok Lee, Xiluo He, Jihwan Lee, Helin Wang, Shrikanth Narayanan,\\
Thomas Thebaud, Laureano Moro-Velazquez, Jes\'us Villalba, and Najim Dehak%
\thanks{Junhyeok Lee, Xiluo He, Helin Wang, Thomas Thebaud, Laureano
Moro-Velazquez, Jes\'us Villalba, and Najim Dehak are with the Center for
Language and Speech Processing, Johns Hopkins University, Baltimore, MD,
USA (e-mail: jlee843@jhu.edu; ndehak3@jhu.edu).}%
\thanks{Jihwan Lee and Shrikanth Narayanan are with the Signal Analysis and
Interpretation Laboratory, University of Southern California, Los Angeles,
CA, USA.}%
\thanks{Junhyeok Lee and Najim Dehak are the corresponding authors.}}

\begin{document}

\maketitle

\begin{abstract}
Neural audio codecs optimized for mel-spectrogram reconstruction often fail to preserve intelligibility. While semantic encoder distillation improves encoded representations, it does not guarantee content preservation in reconstructed speech. In this work, we demonstrate that self-supervised representation reconstruction (SSRR) loss fundamentally improves codec training and performance. First, SSRR significantly accelerates convergence, enabling competitive results after 300k training steps on a single H200 GPU. Second, it enhances intelligibility by reconstructing distilled self-supervised representations from codec outputs. Third, SSRR enables high intelligibility without additional lookahead in streaming Transformer-based codecs, allowing a zero-lookahead architecture for real-time deployment. On LibriSpeech test-clean, JHCodec achieves the best WER and CER among the evaluated codecs while maintaining zero lookahead and low end-to-end latency. We open-source the full implementation, training pipeline, and demo on GitHub\footnote{\url{https://github.com/jhcodec843/jhcodec}}.
\end{abstract}

\begin{IEEEkeywords}
Neural audio codec, self-supervised representation, streaming model.
\end{IEEEkeywords}

\section{Introduction}

\IEEEPARstart{T}{he} rapid evolution of audio and speech large language models \cite{valle,speartts, musicgen,audiolm,moshi} has redefined speech synthesis as a scalable autoregressive language modeling problem.
This paradigm is fundamentally built on speech representations, especially based on neural audio codecs, which serve as tokenizers to compress high-dimensional continuous waveforms into sequences of discrete tokens \cite{soundstream, encodec, dac}.
These codecs typically employ vector-quantized variational autoencoders (VQ-VAE) \cite{vqvae}, with various design choices ranging from single large codebooks \cite{wavtokenizer,bigcodec,ts3codec,magicodec} to multi-stage residual vector quantization (RVQ) \cite{moshi,soundstream,encodec,dac}, alongside alternative approaches like finite scalar quantization (FSQ) \cite{stablecodec} or binary representations \cite{focal,focalstream}.
They are mainly trained with reconstruction objectives, such as multi-scale mel-spectrogram losses, often combined with waveform losses or generative adversarial networks (GANs) \cite{gan} following prior neural vocoders \cite{melgan,hifigan,phaseaug}, to improve perceptual quality.

\begin{figure*}
    \centering
    \includegraphics[width=0.95\linewidth]{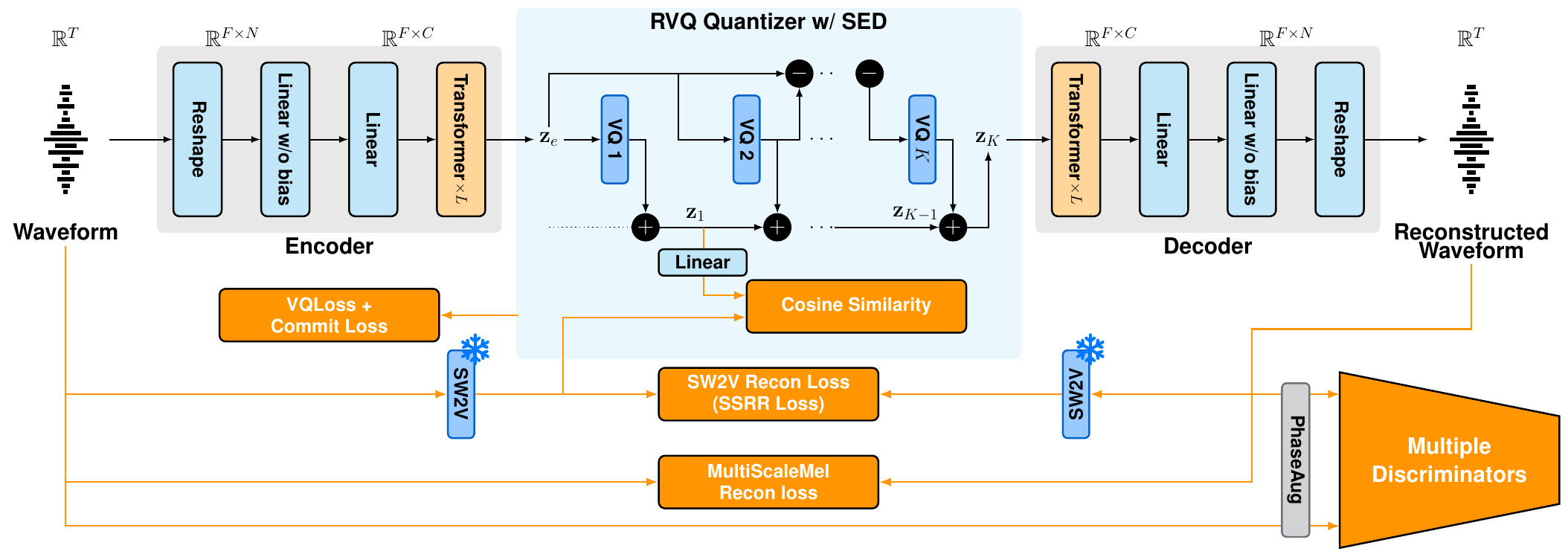}
    \caption{Overall system architecture of JHCodec}
    \label{fig:overall}
\end{figure*}

However, a critical discrepancy has emerged between the primary optimization objectives of these codecs and their application in semantic generation tasks.
When representations optimized solely for acoustic fidelity are applied for semantic tasks, they may exhibit semantic deficiencies \cite{xcodec} that compromise linguistic preservation, and may also be sensitive to perturbations \cite{inconsistency}, such as temporal slicing and phase perturbation \cite{phaseaug}.
To address this, recent research has focused on integrating auxiliary losses into the \textbf{codec encoder} to enforce discrete representation consistency during training.
One of the most prominent approaches is \textbf{semantic encoder distillation} (SED) \cite{moshi,xcodec,speechtokenizer}\footnote{We acknowledge the controversy surrounding the terminology `semantic' \cite{ss_phonetic}, but we use it for consistency with previous literature.}, which aligns the codec's quantized representations with those of a \textbf{self-supervised representation} (SSR) learning model \cite{w2v2,hubert,wavlm, w2vb2}.
Another method involves feeding an additional consistency loss to the codec encoder, enforcing consistency through perceptually invariant augmentations such as slicing or phase augmentation \cite{inconsistency, phaseaug}.
These distillation and auxiliary-loss approaches demonstrate superior generative performance compared to acoustic-only-trained codecs \cite{moshi,xcodec,inconsistency}.
However, these semantic encoder distillation methods do not guarantee the intelligibility or semantic consistency of the decoder's output.
A key reason is that these methods focus solely on the encoder and impose no loss on the decoder, thereby failing to ensure the intelligibility of the reconstructed audio.
Several papers have already highlighted a semantic-acoustic conflict \cite{moshi,xytokenizer}, noting that codecs trained with semantic distillation often suffer from acoustic quality issues, particularly at low bitrates.
Critically, low-bps models are typically evaluated against bitrate-constrained acoustic models without reporting intelligibility objective metrics such as WER.

Rather than using SED, SSR can be treated directly as a reconstruction target, similar to the mel-spectrogram, and it consists of differentiable modules.
We refer to this objective as the \textbf{self-supervised representation reconstruction} (SSRR).
While conceptually similar to perceptual loss functions in the image domain \cite{perceptual} and speech enhancement \cite{perceptualse}, the only prior work that applies this strategy in neural audio codecs is TAAE \cite{stablecodec}.
However, TAAE applies SSRR only at the final stage of training and reports relatively low intelligibility. 
Moreover, it does not provide clear evidence on how SSRR contributes to improving training dynamics.
In addition, due to relatively large model size and CPU-bound operations, recent models, such as W2V-BERT 2.0 \cite{w2vb2}, are not commonly used. 

The emergence of speech-to-speech models \cite{moshi, hibiki} necessitates fully streaming codecs for real-time applications.
While existing streaming models \cite{moshi,encodec,ts3codec,magicodec,focalstream} have achieved competitive results, some rely on a large frame size \cite{moshi}, while others require lookahead mechanisms \cite{magicodec,focalstream} to maintain quality, thereby compromising low-latency requirements.
Even with causal distillation techniques, streaming models often exhibit lower intelligibility than their non-streaming counterparts \cite{focalstream}.
Architectures like TS3-Codec \cite{ts3codec}, which use Transformer-only designs, offer low computational cost but suffer from limited bitrates and semantics-less training, leading to low intelligibility.
MagiCodec \cite{magicodec} attempts to mitigate this with multi-stage training and masking, but it still faces similar challenges.
It is also worth noting that some approaches apply streaming only to the decoder \cite{nanocodec}, but such codecs are not fully streamable, limiting their application to real-time speech-to-speech models.

In this work, we propose \textbf{JHCodec}, a streaming Transformer-based neural audio codec that prioritizes high-intelligibility reconstruction under strict low-latency constraints.
Although recent codecs emphasize performance on downstream generative tasks, evidence from the image domain suggests that downstream quality is not always strongly correlated with reconstruction quality \cite{rvg1, rvg2, rvg3}.
At the same time, prior studies indicate that pure reconstruction quality can serve as an upper bound for generation, with generation quality being further improved by training details \cite{heptapod}.
Motivated by this perspective, we focus on speech reconstruction quality, particularly intelligibility, while ensuring low latency operation for practical speech applications.
To this end, we adopt a high-bitrate, zero-lookahead architecture and introduce the SSRR loss to guide optimization toward linguistically meaningful representations rather than purely mel-spectrogram reconstruction. 
Unlike prior work that applies similar losses only in the later training stage and primarily evaluates signal-level metrics, we systematically study the effect of SSRR from early training across multiple RVQ configurations, with a focus on intelligibility (WER).

Our experiments show that SSRR substantially accelerates codec training, particularly during the early stages, while consistently improving all speech-related metrics.
Notably, SSRR leads to significant gains in intelligibility under all bitrate settings by redefining the optimization objective toward linguistically meaningful representations rather than the mel-spectrogram.
Moreover, incorporating SSRR enables efficient training with only one or two GPUs while achieving performance competitive with state-of-the-art baselines trained with large-scale, multi-node budgets.
Overall, these results demonstrate that SSRR is an effective and practical component for neural audio codecs, enabling our \textbf{JHCodec} to achieve strong reconstruction performance under strict low-latency constraints.

\section{Method}

\subsection{Model Architecture}
We adopt the fully causal Transformer architecture, inspired by TS3-Codec \cite{ts3codec}, accelerated by FlashAttention~\cite{flashattn, flashattn2} optimized for low latency.
TS3-Codec demonstrates high computational efficiency with a low number of multiply–accumulate operations (MAC), despite its large parameter count.

Our architecture is built upon TS3-Codec, replacing its single-codebook VQ with RVQ to overcome the limited intelligibility imposed by its single-codebook VQ's limited representational capacity.
We also incorporate modern Transformer design principles, including Pre-Layer Normalization (PreLN) \cite{preln}, rotary positional embeddings \cite{rotary}, SwiGLU activation for feed-forward networks \cite{glu}, and LayerScale \cite{layerscale}.
To enhance training stability, we retain LayerNorm \cite{layernorm} instead of replacing it with RMSNorm \cite{rmsnorm}.
Figure \ref{fig:overall} illustrates the overall architecture and the applied losses.

Specifically, our model utilizes an $N=320$-sample window for input reshaping.
This representation is then sequentially upsampled to 768 and subsequently to $C=1024$ dimensions using two linear layers.
The encoder and decoder components each comprise $L=8$ Transformer decoder layers.
Within these Transformer layers, we use a model dimension of 1024, which is expanded to 4096 in the feed-forward network (FFN).
Additionally, all sliding window sizes are reduced to 16.

Reducing the frame rate in neural audio codecs improves computational efficiency but introduces a trade-off with intelligibility, as observed in TS3-Codec \cite{ts3codec} and other prior studies \cite{flexicodec}.
To compensate for degraded intelligibility at low frame rates, such as 12.5 Hz, recent state-of-the-art codecs adopt deep RVQ hierarchies. 
For instance, Mimi \cite{moshi} employs 32 codebooks. 
However, combining a low frame rate with deep RVQ introduces two critical issues.
First, a lower frame rate increases overall system latency, as each frame spans a longer temporal interval, thereby increasing the latency before decoding can proceed.
Second, deep RVQ hierarchies significantly increase computational cost and latency due to repeated in-and-out projections across multiple quantization stages. 
These sequential quantization steps cannot be fully parallelized, further exacerbating the efficiency overhead.
Moreover, for downstream speech-to-speech applications, Mimi is commonly configured with only 8 codebooks \cite{moshi}, since using all 32 results in substantial computational overhead.
Therefore, we select a 50 Hz high frame rate configuration with $K=8$ codebooks to achieve high intelligibility while maintaining low latency.


To improve the system's overall computational efficiency, we applied FlashAttention \cite{flashattn,flashattn2} to all attention layers.
Notably, while some prior models do not provide an official streaming implementation, our model supports efficient streaming inference via KV caching \cite{kvcache}.

\subsection{Self-Supervised Representation}\label{section:ssr}
Previous studies have indicated that generating audio from semantic representations can enhance intelligibility \cite{audiolm,speechtokenizer}.
Following this, Mimi \cite{moshi} explored distilling semantic information from WavLM \cite{wavlm} into the first VQ codebooks via cosine similarity.
Without lookahead mechanisms, causal representations exhibit a high phoneme error rate, suggesting a distinct pattern compared to non-causal models \cite{effectivecontext}.
Consequently, our objective is to extract a reliable, causal, and lightweight self-supervised speech representation.

To achieve this, we introduce an explicit model that causally distills self-supervised representations, and we refer to this model as \textbf{SW2V}.
Similar to Mimi's SED approach \cite{moshi}, SW2V is trained to maximize the cosine similarity with the original self-supervised model's representations.
We choose multilingually trained W2V-BERT 2.0\footnote{\url{https://hf.co/facebook/w2v-bert-2.0}} \cite{w2vb2} for potential future multilingual extensions, as WavLM \cite{wavlm} is trained only on English datasets, which may limit multilingual generalization.
Consistent with prior work \cite{xcodec2}, we utilize features from the 17th layer of W2V-BERT 2.0.
In addition, since W2V-BERT 2.0 also uses 1024-dimensional embeddings, identical to our model's dimensions, no additional layers are required to align its representation dimensionality with our model.
SW2V shares the same architectural design as our codec's encoder to achieve both efficient calculation and causal architecture. 

\subsection{RVQ-VAE Neural Audio Codec}
We adopt a neural audio compression framework based on the residual vector quantized variational auto-encoder (RVQ-VAE), drawing inspiration from prominent codecs such as DAC \cite{dac} and Mimi \cite{moshi}.
The model architecture consists of an encoder $E$, a quantizer $Q$, and a decoder $G$.
Given a raw audio waveform $\mathbf{x} \in \mathbb{R}^{T}$, the encoder maps it to a sequence of continuous latent representations $\mathbf{z}_e = E(\mathbf{x}) \in \mathbb{R}^{F \times C}$, where $T$ is the number of audio samples, $F$ represents the number of frames, and $C$ denotes the embedding channel dimension.
Since quantization is applied independently to each frame in the $F$-length sequences, we simply describe the formulation at the frame level for clarity. 
For brevity, we omit the frame index and denote $\mathbf{z}_e \in \mathbb{R}^C$ in the following derivations.

To quantize the continuous latent representations $\mathbf{z}_e$ from the encoder, we employ RVQ.
The quantizer $Q$ comprises a sequence of $K$ vector quantizers $\{q_1, \dots, q_K\}$.
The quantization process is performed iteratively, where the input of the $i$-th quantizer is the residual error from the preceding stages, denoted as $\mathbf{r}_i$, and the corresponding quantized embedding is $\Tilde{\mathbf{z}}_i$, defined as: 
\begin{equation}
\mathbf{r}_1 = \mathbf{z}_e,\, \mathbf{r}_i=\mathbf{r}_{i-1} -  \Tilde{\mathbf{z}}_{i-1}=\mathbf{z}_e - \sum_{j=1}^{i-1} \Tilde{\mathbf{z}}_j \quad \text{for }\,2\le i\le K.
\end{equation}
To enhance robustness and enable variable bitrate operation, we adopt quantization dropout, proposed by Kumar \etal \cite{dac}, in which only the first $k$ quantizers are used, with $1 \le k \le K$.
Under this quantization dropout scheme, the quantized representation using the first $k$ quantizers is defined as:
\begin{equation}
    \mathbf{z}_k = Q(\mathbf{z}_e) = \sum_{i=1}^{k} \Tilde{\mathbf{z}}_i, \quad \text{where} \quad \Tilde{\mathbf{z}}_i = q_i(\mathbf{r}_i).
\end{equation}
Furthermore, Kumar \etal \cite{dac} advocate for the use of input and output projections \cite{vqgan} during residual quantization to increase the codebook utilization, expressed as:
\begin{equation}
q_i(\mathbf{r}_i) = \mathbf{W}_{i, \mathrm{out}} \mathbf{e}_{i,v_i}, \, \text{ where } \, v_i = \underset{j}{\mathrm{argmin}}\lVert \mathbf{W}_{i,\mathrm{in}}\mathbf{r}_i - \mathbf{e}_{i,j} \rVert^2_2.
\end{equation}
Here, the $i$-th vector quantizer consists of an input projection $\mathbf{W}_{i,\mathrm{in}} \in \mathbb{R}^{M\times C}$, an output projection $\mathbf{W}_{i,\mathrm{out}} \in \mathbb{R}^{C\times M}$, the $i$-th VQ quantizer with codebooks $\{\mathbf{e}_{i,1},\mathbf{e}_{i,2},\dotsc,\mathbf{e}_{i,V} \}$, and the closest codebook index $v_i \in\{1,2,\dotsc,V\}$.
In this formulation, $M$ is significantly smaller than $C$ to achieve low-rank compression, and $V$ denotes the codebook's vocabulary size.
We set $M=16$, following the low-dimensional configuration of TS3-Codec \cite{ts3codec}.
In addition, we use the Euclidean distance VQ, rather than the cosine similarity VQ.

During VQ, to enable gradient propagation through a vector quantizer, we employ the straight-through estimation (STE) \cite{vqvae}.
When input and output projection layers are introduced \cite{vqgan}, the gradient flow through ${q}_i$ is given by:
\begin{align}
   \frac{\partial q_i(\mathbf{r}_{i})}{\partial \mathbf{r}_i}  &=  \frac{\partial \Tilde{\mathbf{z}}_i}{\partial \mathbf{r}_i} =   \frac{\partial (\mathbf{W}_{i,\mathrm{out}} \mathbf{e}_{i,v_i})}{\partial \mathbf{e}_{i,v_i}} \nonumber  \frac{\partial \mathbf{e}_{i,v_i}}{\partial (\mathbf{W}_{i,\mathrm{in}}\mathbf{r}_i)} \frac{\partial (\mathbf{W}_{i,\mathrm{in}}\mathbf{r}_i)}{\partial \mathbf{r}_i}  \\
   &=   \mathbf{W}_{i,\mathrm{out}}   \frac{\partial \mathbf{e}_{i,v_i}}{\partial (\mathbf{W}_{i,\mathrm{in}}\mathbf{r}_i)} \mathbf{W}_{i,\mathrm{in}} \approx \mathbf{W}_{i,\mathrm{out}} \mathbf{W}_{i,\mathrm{in}},
\end{align}
where $\frac{\partial \mathbf{e}_{i,v_i}}{\partial (\mathbf{W}_{i,\mathrm{in}}\mathbf{r}_i)}$ is approximated bt the identity matrix $\mathbf{I}$ through the STE.
Therefore, the gradient through the quantizer $Q$ to the continuous latent, including quantizer drop, can be represented as follows:
\begin{align}
\label{eq:rvq_grad}
  & \frac{\partial \mathbf{z}_k}{\partial \mathbf{z}_e}  = \frac{\partial\left(\sum_{i=1}^k \Tilde{\mathbf{z}}_i \right)}{\partial\mathbf{z}_e} = \sum_{i=1}^k \frac{\partial\Tilde{\mathbf{z}}_i}{{\partial\mathbf{r}_i}} \frac{{\partial\mathbf{r}_i}}{{\partial\mathbf{z}_e}} 
   \nonumber \\
   & \approx   \Big( \mathbf{W}_{1,\mathrm{out}} \mathbf{W}_{1,\mathrm{in}}  \nonumber \\ 
   &  \phantom{\approx\frac{\partial \mathcal{L}}{\partial \mathbf{z}_k} \Big(}  + \mathbf{W}_{2,\mathrm{out}} \mathbf{W}_{2,\mathrm{in}}(\mathbf{I} - \mathbf{W}_{1,\mathrm{out}} \mathbf{W}_{1,\mathrm{in}}) + \cdots \Big)  \nonumber \\
   & =  \left( \sum_{i=1}^k \mathbf{W}_{i,\mathrm{out}} \mathbf{W}_{i,\mathrm{in}} \left( \prod_{\substack{j=1, \\ \mathrm{if} \,i>1 \\\mathrm{else} \,\mathbf{I}}}^{i-1} \left(\mathbf{I} -\mathbf{W}_{j,\mathrm{out}} \mathbf{W}_{j,\mathrm{in}} \right)\right)\right).
\end{align}
We backpropagate gradients through the encoder only via the quantized embeddings, unlike Mimi \cite{moshi}, which uses unquantized embeddings.

In addition, the RVQ module is trained using the standard VQ loss and commitment loss to jointly update the encoder and the codebooks.
Specifically, we adopt the loss formulation from VQ-VAE \cite{vqvae}, which encourages the encoder outputs to commit to discrete codebook entries while allowing the codebooks to adapt to the data distribution.
Both losses are as follows:
\begin{align}
    \mathcal{L}_\mathrm{vq}  &=\sum_{i=1}^{k}\lVert \mathrm{sg}[\mathbf{W}_{i,\mathrm{in}}\mathbf{r}_i] - \mathbf{e}_{i,v_i} \rVert^2_2, \\
    \mathcal{L}_\mathrm{commit}  &=\sum_{i=1}^{k}\lVert \mathbf{W}_{i,\mathrm{in}}\mathbf{r}_i - \mathrm{sg}[\mathbf{e}_{i,v_i}] \rVert^2_2,
\end{align}
where $\mathrm{sg}$ refers to the stop gradient operation.

Mimi \cite{moshi} integrates SED into DAC's RVQ, employing two types of codebooks: acoustic and semantic.
The acoustic codebooks operate identically to DAC's RVQ.
The semantic codebook utilizes only one VQ layer and applies a cosine similarity loss derived from a self-supervised model.
The final quantized embedding is obtained by summing the outputs of these two components.
While the original Mimi used WavLM \cite{wavlm}, we use a distilled causal SW2V, as described in section \ref{section:ssr}.
Furthermore, unlike direct WavLM distillation, which would require learning causal inference from a bidirectional model, SW2V inherently serves as an upper limit for a causal model's performance.
In particular, since the semantic representations from the semantic codec are quantized, they cannot match the performance of the continuously trained SW2V.

To compare RVQ setups with and without SED in the TS3-Codec-based Transformer architecture, we trained both models with identical configurations, differing only in the RVQ setup.
For both RVQ styles, we utilize a vocabulary size of $V=1024$ for each codebook, and we employ $K=8$ codebooks.

The resulting quantized embedding $\mathbf{z}_k$ is then fed into the decoder to synthesize the reconstructed waveform $\hat{\mathbf{x}} = G(\mathbf{z}_k)$.
The model is trained with a comprehensive objective function that includes a multi-scale mel reconstruction loss with L1 distance $\mathcal{L}_\mathrm{mel}$ \cite{musicgen}, adversarial losses $\mathcal{L}_\mathrm{adv}$, and feature-matching losses $\mathcal{L}_\mathrm{fm}$ derived from discriminators.
We utilize the multiple discriminators following MPD \cite{hifigan} and MS-STFTD \cite{encodec}.
In addition, phase perturbation has already shown effectiveness in both vocoder \cite{phaseaug} and codec training \cite{inconsistency}. 
We apply PhaseAug\footnote{\url{https://github.com/maum-ai/phaseaug}} as a differentiable GAN augmentation.
The overall training objective follows the standard loss formulation for the neural audio codec is expressed as:
\begin{align}
    \mathcal{L}_\mathrm{codec} = &\lambda_\mathrm{mel} \mathcal{L}_\mathrm{mel} +  \lambda_\mathrm{vq} \mathcal{L}_\mathrm{vq} +  \lambda_\mathrm{commit}\mathcal{L}_\mathrm{commit} \nonumber \\
    & + \lambda_\mathrm{adv}\mathcal{L}_\mathrm{adv} + \lambda_\mathrm{fm}\mathcal{L}_\mathrm{fm},
\end{align}
where $\lambda_\mathrm{mel}$, $\lambda_\mathrm{vq}$, $\lambda_\mathrm{commit}$, $\lambda_\mathrm{adv}$ and $\lambda_\mathrm{fm}$ are loss coefficients.
The overall model is illustrated in Figure \ref{fig:overall}.

To make our codec more robust to noise, we add a small amount of noise to the encoder input, as GAN-based training has been shown to enable implicit upsampling~\cite{moshi} and denoising~\cite{soundstream}.
For 10\% of the training batches, we randomly add either Gaussian or sinusoidal noise to encourage the model to implicitly learn to denoise stationary noise.

\subsection{Self-Supervised Representation Reconstruction Loss}
Despite the impressive reconstruction results of neural audio codecs, extensive research indicates that VQ often degrades intelligibility and speaker similarity compared to continuous features \cite{pits,vevo,maskvct},  particularly in streaming models.
One reason is that, in the current loss $\mathcal{L}_\mathrm{codec}$, intelligibility is influenced only indirectly through the multi-scale mel-spectrogram reconstruction loss and the feature-matching losses.
Although a zero-valued loss would trivially correspond to identical intelligibility, the loss magnitude does not directly reflect the perceptual or linguistic intelligibility of the output speech.

To mitigate this limitation in the loss of the neural audio codec, 
we introduce the \textbf{self-supervised representation reconstruction (SSRR) loss} ($\mathcal{L}_{ssrr}$), a more intuitive proxy for intelligibility based on the distance between self-supervised representations, explicitly capturing linguistic consistency beyond low-level acoustic similarity.
We set the SW2V as the target representation.  
This loss quantifies and penalizes the semantic discrepancy between the original audio $\mathbf{x}$ and the reconstructed audio $\hat{\mathbf{x}}$:
\begin{equation}
    \mathcal{L}_\mathrm{ssrr} = \lVert \Phi(\mathbf{x}) - \Phi(\hat{\mathbf{x}}) \rVert_1.
\end{equation}
Here, $\Phi(\cdot)$ represents the operation of extracting features from the frozen SW2V.
We adopt the L1 loss rather than the cosine similarity loss from SED, as the existing perceptual losses \cite{perceptual,perceptualse,stablecodec} and feature-matching losses \cite{hifigan} are known to be effective with the L1 or L2 loss, despite their target feature extractors being trained on different objectives.

By minimizing this objective, the gradient propagates backward through the decoder $G$, the quantizer $Q$, and the encoder $E$, compelling the codec to retain the phonetic information necessary to accurately reconstruct the SW2V features. 
While the standard GAN ($\mathcal{L}_\mathrm{adv}$ and $\mathcal{L}_\mathrm{fm}$) and multi-scale mel reconstruction losses ($\mathcal{L}_\mathrm{mel}$) do not explicitly guarantee the preservation of phonetic content under the quantizer drop, 
$\mathcal{L}_\mathrm{ssrr}$ explicitly enforces the retention of phonetic information, thereby improving intelligibility.
The total training objective is as follows:
\begin{equation}
    \mathcal{L}_\mathrm{total} =  \mathcal{L}_\mathrm{codec}  + \lambda_\mathrm{ssrr}\mathcal{L}_\mathrm{ssrr}.
\end{equation}

\section{Experiments}
\subsection{Training Details}
For all experiments, we used the AdamW optimizer \cite{adamw} with a learning rate of $1\times10^{-4}$ and a weight decay of $1\times10^{-2}$.
All audio samples were resampled to 16 kHz.
During training, utterances from the same speaker were concatenated to form fixed-length inputs of 10.24 seconds.
All training runs used one H200 GPU, except for JHCodec-8 after 600k steps, which used two H200 GPUs.
All batch sizes were set to the maximum values that fully utilize the available GPU memory.

During SW2V training, we used a batch size of 300 and trained for 60k steps, stopping before instability occurred.
The resulting SW2V model achieved an average cosine similarity of 0.9 or higher on the LibriTTS-R development set.

A batch size of 42 was used during codec training.
We set $\lambda_\mathrm{mel}=0.1$, $\lambda_\mathrm{vq} =1$, $\lambda_\mathrm{commit} =0.1$, $\lambda_\mathrm{adv} =1$, $\lambda_\mathrm{fm} =1$.
The SSRR weight was set to $\lambda_\mathrm{ssrr} =1$ when SSRR was enabled, and $\lambda_\mathrm{ssrr} =0$ otherwise.
For each batch, we tracked the vocabulary usage of every codebook using an exponential moving average (EMA) with a decay rate of 0.99.
If the EMA usage of a codebook entry fell below 0.90, we expired and reinitialized that entry with a randomly selected vector from the current batch.
For the first 10k steps, the model was trained without GAN objectives and without the SSRR loss.
We empirically found that jointly training with these components hinders training stability in the initial stages.
From 10k, GAN training and the SSRR loss were enabled, and we applied masking by replacing 10\% of both encoder and decoder inputs with special mask tokens.
After 100k steps, the masking was removed, and training continued with the full set of objectives.
Beyond 600k steps, we used two H200 GPUs and trained the model to 1.4M steps, which corresponds to 2.2M steps under a single-GPU setting.
All other details are \textbf{already open-sourced}.

Unlike the typical GAN training, we updated the generator before the discriminator as G$\rightarrow$D to improve training stability.
Using the conventional D$\rightarrow$G update scheme led to unstable training on noisy datasets.
Following VITS \cite{vits}, we reduced the memory footprint of the discriminators by randomly slicing 2.56-second audio segments as inputs to the discriminators.
Notably, the discriminators consumed significantly more GPU memory than the codec during training.
It could also be observed that the batch size used for training SW2V is substantially larger than that used for the codec, even though SW2V consumes only about half the GPU memory.

\subsection{Datasets}
Our models were trained on diverse corpora to enhance its generalization for English varieties and beyond. 
We utilized the train-clean subsets of LibriTTS‑R \cite{librittsr}, the train subset of MLS‑en \cite{mls}, VCTK \cite{vctk}, LibriHeavy-Large \cite{librilight,libriheavy}, the clean subset of HiFi‑TTS \cite{hifitts}, LJSpeech \cite{ljspeech}, the speech subset of RAVDESS \cite{ravdess}, and the English subset of Emilia \cite{emilia}.
We employed dataset balancing to increase the sampling probability of clean speech data. 

For evaluation, we used the LibriSpeech \cite{librispeech} test-clean subset for clean speech and the test-other subset for noisy speech.
To further assess robustness under extreme noise conditions, we also evaluated on the TITW-Hard test set \cite{titw}.
We also assessed its generalization to multilingual scenarios by testing on the MLS's non-English test sets \cite{mls}, including Dutch, French, German, Italian, Polish, Portuguese, and Spanish.

\subsection{Metrics}
Existing benchmarks \cite{codecsuperb,codecbench} support a wide range of downstream tasks and signal reconstruction metrics, but does not directly assess whether the core speech content is preserved during reconstruction.
We also measured STOI \cite{stoi}, and it does not reflect intelligibility for similar but failed reconstruction.
Therefore, we focus on metrics that are more relevant to speech synthesis.
We report word error rate (WER) and character error rate (CER) as automatic speech recognition (ASR)–based proxy metrics for intelligibility, measured by Whisper Large-v3\footnote{\url{https://hf.co/openai/whisper-large-v3}} \cite{whisper}.
Speaker similarity (S-SIM) is measured as the cosine similarity between WavLM speaker embeddings\footnote{\url{https://hf.co/microsoft/wavlm-base-plus-sv}} \cite{wavlm} extracted from the original and reconstructed speech signals.
Perceptual speech quality is evaluated using the UTMOS v2 \cite{utmosv2}.
For the TITW-Hard test set, we compare differential WER (dWER), i.e., the difference between the original and reconstructed speech transcriptions, as this dataset only contains transcripts from an outdated version of Whisper rather than ground-truth transcriptions.

\subsection{Ablations}

We conduct ablation studies on the LibriSpeech test-clean set to analyze (1) the impact of RVQ design choices and (2) the effect of SSRR loss, using models trained for 300k and 600k steps.
The suffix D and M indicate DAC-style and Mimi-style RVQ configurations, respectively.
The numeric suffix denotes the number of codebooks utilized in the inference.
Results at the maximum bitrate are summarized in Table~\ref{tab:results_ablation}, while results across all bitrates are illustrated in Figure \ref{fig:full_ablation}.

\begin{table}[h]
    \caption{
    Evaluation results for JHCodec-8 ablation study.
    }
    \label{tab:results_ablation}
    \centering
    \resizebox{0.99\linewidth}{!}{%
    \begin{tabular}{l|cc|cc|cc}
    \toprule
    Steps
     & SED
     & SSRR
     & WER ($\downarrow$)  
     & CER ($\downarrow$) 
     & S-SIM ($\uparrow$)  
     & UTMOS ($\uparrow$) 
       \\
    \midrule
    Ground Truth (GT)&
    $-$ & $-$ & 2.99  & 1.13 &  1.0000 & 3.2311
     \\      
    \midrule
    300k & \xmark & \xmark & 6.28  & 3.15 & 0.9287 & \textbf{3.3143} \\
    300k & \xmark & \cmark  & \textbf{3.54}  & \textbf{1.38} & \textbf{0.9631} & 3.2100 \\
    \midrule
    300k & \cmark  & \xmark  & 5.43 & 2.48 & 0.9290 & \textbf{3.2030} \\
    300k & \cmark &\cmark  & \textbf{3.57} & \textbf{1.50} & \textbf{0.9698} & 3.1712 \\
    \midrule
    600k & \xmark & \xmark & 4.42  & 1.88 & 0.9415 & \textbf{3.2790} \\
    600k & \xmark & \cmark  & \textbf{3.31}  &  \textbf{1.31}  & \textbf{0.9759} & {3.2663}  \\
    \midrule
    600k & \cmark & \xmark  & 4.69  &  2.07  & 0.9406 & \textbf{3.2851}  \\
    600k & \cmark &\cmark  & \textbf{3.29} & \textbf{1.29} & \textbf{0.9783} & 3.1697  \\
    \midrule
    1.0M & \cmark &\cmark &  \textbf{3.19}   &  \textbf{1.25} & \textbf{0.9826} &  \textbf{3.3229} \\
    \midrule
    1.4M & \cmark & \cmark & \textbf{3.04} & \textbf{1.13} & \textbf{0.9855} & \textbf{3.3167} \\
    \bottomrule
    \end{tabular}
    }
\end{table}

\pgfplotsset{
    rowstyle/.style={
        width=6.05cm, 
        height=4.45cm,
        grid=both,
        major grid style={line width=.2pt,draw=gray!50},
        title style={yshift=-2pt, font=\bfseries\small},
        tick label style={font=\scriptsize},
        label style={font=\scriptsize},
        line width=1pt,
        mark size=1.5pt
    }
}
\begin{figure*}[htbp]
    \centering
    
    \begin{minipage}{0.33\textwidth}
        \centering
        \begin{tikzpicture}
            \begin{axis}[rowstyle, ylabel={WER (\%)}, title={WER (300k)}, xmin=0.25, xmax=4.25, ymin=2.5, ymax=15,
                legend pos=north east, legend style={font=\tiny, fill opacity=0.8, row sep=-3pt}]
                \addplot[iblue, line width=1pt] coordinates {(0.0, 2.99) (4.5, 2.99)}; \addlegendentry{Ground Truth}
                \addplot[iorange, mark=*] coordinates {(0.5, 86.37) (1.0, 24.31) (1.5, 12.19) (2.0, 8.65) (2.5, 7.17) (3.0, 6.38) (3.5, 5.94) (4.0, 6.28)}; \addlegendentry{w/o SED w/o SSRR}
                \addplot[igreen, mark=square*] coordinates {(0.5, 24.64) (1.0, 13.93) (1.5, 10.25) (2.0, 8.18) (2.5, 7.09) (3.0, 6.19) (3.5, 5.83) (4.0, 5.43)}; \addlegendentry{w/\phantom{o} SED w/o SSRR}
                \addplot[iorange, mark=o] coordinates {(0.5, 23.83) (1.0, 6.81) (1.5, 4.65) (2.0, 4.02) (2.5, 3.83) (3.0, 3.68) (3.5, 3.55) (4.0, 3.54)}; \addlegendentry{w/o SED w/\phantom{o} SSRR}
                \addplot[igreen, mark=square] coordinates {(0.5, 7.81) (1.0, 5.79) (1.5, 4.64) (2.0, 4.01) (2.5, 3.80) (3.0, 3.62) (3.5, 3.54) (4.0, 3.57)}; \addlegendentry{w\phantom{o} SED w/\phantom{o} SSRR}
            \end{axis}
        \end{tikzpicture}
    \end{minipage}\hfill
    \begin{minipage}{0.33\textwidth}
        \centering
        \begin{tikzpicture}
            \begin{axis}[rowstyle, 
            ylabel={S-SIM}, title={S-SIM (300k)}, xmin=0.25, xmax=4.25, ymin=0.75, ymax=1.01]
                \addplot[iblue, line width=1pt] coordinates {(0.0, 1.0) (4.5, 1.0)};
                \addplot[iorange, mark=*] coordinates {(0.5, 0.7856) (1.0, 0.8581) (1.5, 0.8928) (2.0, 0.9089) (2.5, 0.9172) (3.0, 0.9231) (3.5, 0.9260) (4.0, 0.9287)};
                \addplot[iorange, mark=o] coordinates {(0.5, 0.8728) (1.0, 0.9224) (1.5, 0.9399) (2.0, 0.9515) (2.5, 0.9564) (3.0, 0.9595) (3.5, 0.9615) (4.0, 0.9631)};
                \addplot[igreen, mark=square*] coordinates {(0.5, 0.8023) (1.0, 0.8799) (1.5, 0.8957) (2.0, 0.9065) (2.5, 0.9138) (3.0, 0.9232)(3.5, 0.9271) (4.0, 0.9290)};
                \addplot[igreen, mark=square] coordinates {(0.5, 0.8971) (1.0, 0.9270) (1.5, 0.9428) (2.0, 0.9534) (2.5, 0.9603)
                (3.0, 0.9652) (3.5, 0.9678) (4.0, 0.9698)};
            \end{axis}
        \end{tikzpicture}
    \end{minipage}\hfill
    \begin{minipage}{0.33\textwidth}
        \centering
        \begin{tikzpicture}
            \begin{axis}[rowstyle, ylabel={UTMOS}, title={UTMOS (300k)}, xmin=0.25, xmax=4.25, ymin=2.0, ymax=3.4]
                \addplot[iblue, line width=1pt] coordinates {(0.0, 3.23) (4.5, 3.23)};
                \addplot[iorange, mark=*] coordinates {(0.5, 2.1204) (1.0, 2.9155) (1.5, 3.1039) (2.0, 3.2072) (2.5, 3.2635) (3.0, 3.2784) (3.5, 3.2979) (4.0, 3.3143)};
                \addplot[iorange, mark=o] coordinates {(0.5, 2.4415) (1.0, 2.9795) (1.5, 3.1129) (2.0, 3.1384) (2.5, 3.1751) (3.0, 3.1835) (3.5, 3.1986) (4.0, 3.2100)};
                \addplot[igreen, mark=square*] coordinates {(0.5, 2.2026) (1.0, 2.7645) (1.5, 2.9271) (2.0, 3.0320) (2.5, 3.1287) (3.0, 3.1754) (3.5, 3.1945)  (4.0, 3.2030)};
                \addplot[igreen, mark=square] coordinates {(0.5, 2.4045) (1.0, 2.9754) (1.5, 3.0031) (2.0, 3.0646) (2.5, 3.1198) (3.0, 3.1416) (3.5, 3.1744) (4.0, 3.1712)};
            \end{axis}
        \end{tikzpicture}
    \end{minipage}

    \begin{minipage}{0.33\textwidth}
        \centering
        \begin{tikzpicture}
            \begin{axis}[rowstyle, ylabel={WER (\%)}, title={WER (600k)}, xmin=0.25, xmax=4.25, ymin=2.5, ymax=12,
                legend pos=north east, legend style={font=\tiny, fill opacity=0.8, row sep=-3pt}]
                \addplot[iblue, line width=1pt] coordinates {(0.0, 2.99) (4.5, 2.99)}; \addlegendentry{Ground Truth}
                \addplot[iorange, mark=*] coordinates {(0.5, 78.86) (1.0, 18.00) (1.5, 8.52) (2.0, 6.70) (2.5, 5.48) (3.0, 4.97) (3.5, 4.67) (4.0, 4.42)}; \addlegendentry{w/o SED w/o SSRR}
                \addplot[igreen, mark=square*] coordinates {(0.5, 18.56) (1.0, 10.96) (1.5, 8.37) (2.0, 6.76) (2.5, 5.84) (3.0, 5.07) (3.5, 4.89) (4.0, 4.69)}; \addlegendentry{w/\phantom{o} SED w/o SSRR}
                \addplot[iorange, mark=o] coordinates {(0.5, 23.84) (1.0, 6.03) (1.5, 4.36) (2.0, 3.79) (2.5, 3.61) (3.0, 3.54) (3.5, 3.43) (4.0, 3.31)}; \addlegendentry{w/o SED w/\phantom{o} SSRR}
                \addplot[igreen, mark=square] coordinates {(0.5, 7.42) (1.0, 5.23) (1.5, 4.26) (2.0, 3.70) (2.5, 3.59) (3.0, 3.44) (3.5, 3.36) (4.0, 3.29)}; \addlegendentry{w/\phantom{o} SED w/\phantom{o} SSRR}
            \end{axis}
        \end{tikzpicture}
    \end{minipage}\hfill
    \begin{minipage}{0.33\textwidth}
        \centering
        \begin{tikzpicture}
            \begin{axis}[rowstyle, ylabel={S-SIM}, title={S-SIM (600k)}, xmin=0.25, xmax=4.25, ymin=0.8, ymax=1.01]
                \addplot[iblue, line width=1pt] coordinates {(0.0, 1.0) (4.5, 1.0)};
                \addplot[iorange, mark=*] coordinates {(0.5, 0.8026) (1.0, 0.8778) (1.5, 0.9085) (2.0, 0.9225) (2.5, 0.9319) (3.0, 0.9368) (3.5, 0.9382) (4.0, 0.9415)};
                \addplot[igreen, mark=square*] coordinates {(0.5, 0.8528) (1.0, 0.8924) (1.5, 0.9075) (2.0, 0.9181) (2.5, 0.9263) (3.0, 0.9342) (3.5, 0.9378) (4.0, 0.9406)};
                \addplot[iorange, mark=o] coordinates {(0.5, 0.8718) (1.0, 0.9366) (1.5, 0.9530) (2.0, 0.9640) (2.5, 0.9689) (3.0, 0.9725) (3.5, 0.9744) (4.0, 0.9759)};
                \addplot[igreen, mark=square] coordinates {(0.5, 0.8994) (1.0, 0.9370) (1.5, 0.9549) (2.0, 0.9661) (2.5, 0.9708) (3.0, 0.9748) (3.5, 0.9768) (4.0, 0.9783)};

            \end{axis}
        \end{tikzpicture}
    \end{minipage}\hfill
    \begin{minipage}{0.33\textwidth}
        \centering
        \begin{tikzpicture}
            \begin{axis}[rowstyle, ylabel={UTMOS}, title={UTMOS (600k)}, xmin=0.25, xmax=4.25, ymin=1.5, ymax=3.4]
                \addplot[iblue, line width=1pt] coordinates {(0.0, 3.23) (4.5, 3.23)};
                \addplot[iorange, mark=*] coordinates {(0.5, 2.1213) (1.0, 2.9340) (1.5, 3.1300) (2.0, 3.2005) (2.5, 3.2347) (3.0, 3.2537) (3.5, 3.2672) (4.0, 3.2790)};
                \addplot[igreen, mark=square*] coordinates {(0.5, 1.7469) (1.0, 3.1132) (1.5, 3.1117) (2.0, 3.1612) (2.5, 3.2137) (3.0, 3.2701) (3.5, 3.2862) (4.0, 3.2851)};
                \addplot[iorange, mark=o] coordinates {(0.5, 2.4572) (1.0, 2.9844) (1.5, 3.0866) (2.0, 3.1615) (2.5, 3.2052) (3.0, 3.2290) (3.5, 3.2378) (4.0, 3.2663)};
                \addplot[igreen, mark=square] coordinates {(0.5, 2.1980) (1.0, 2.9578) (1.5, 2.9920) (2.0, 3.0626) (2.5, 3.1115) (3.0, 3.1436) (3.5, 3.1582) (4.0, 3.1697)};
            \end{axis}
        \end{tikzpicture}
    \end{minipage}

    \begin{minipage}{0.33\textwidth}
        \centering
        \begin{tikzpicture}
            \begin{axis}[rowstyle, ylabel={WER (\%)}, title={WER (w/ SED w/ SSRR)}, xmin=0.25, xmax=4.25, ymin=2.5, ymax=8,xlabel={Bitrate (kbps)},
                legend pos=north east, legend style={font=\tiny, fill opacity=0.8, row sep=-3pt}]
                \addplot[iblue, line width=1pt] coordinates {(0.0, 2.99) (4.5, 2.99)}; \addlegendentry{Ground Truth}
                \addplot[igreen, mark=square] coordinates {(0.5, 7.81) (1.0, 5.79) (1.5, 4.64) (2.0, 4.01) (2.5, 3.80) (3.0, 3.62) (3.5, 3.54) (4.0, 3.57)}; \addlegendentry{300k}
                \addplot[iorange, mark=o] coordinates {(0.5, 7.42) (1.0, 5.23) (1.5, 4.26) (2.0, 3.70) (2.5, 3.59) (3.0, 3.44) (3.5, 3.36) (4.0, 3.29)}; \addlegendentry{600k}
                \addplot[icyan, mark=triangle*] coordinates {(0.5, 6.64) (1.0, 4.96) (1.5, 4.11) (2.0, 3.65) (2.5, 3.37) (3.0, 3.27) (3.5, 3.19) (4.0, 3.19)}; \addlegendentry{1M}
                \addplot[ired, mark=diamond*] coordinates {(0.5, 6.09) (1.0, 4.59) (1.5, 3.82) (2.0, 3.41) (2.5, 3.29) (3.0, 3.19) (3.5, 3.14) (4.0, 3.04)}; \addlegendentry{1.4M}
            \end{axis}
        \end{tikzpicture}
    \end{minipage}\hfill
    \begin{minipage}{0.33\textwidth}
        \centering
        \begin{tikzpicture}
            \begin{axis}[rowstyle, ylabel={S-SIM}, title={S-SIM (w/ SED w/ SSRR)}, xlabel={Bitrate (kbps)}, xmin=0.25, xmax=4.25, ymin=0.85, ymax=1.01]
                \addplot[iblue, line width=1pt] coordinates {(0.0, 1.0) (4.5, 1.0)};
                \addplot[igreen, mark=square] coordinates {(0.5, 0.8971) (1.0, 0.9270) (1.5, 0.9428) (2.0, 0.9534) (2.5, 0.9603) (3.0, 0.9652) (3.5, 0.9678) (4.0, 0.9698)};
                \addplot[iorange, mark=o] coordinates {(0.5, 0.8994) (1.0, 0.9370) (1.5, 0.9549) (2.0, 0.9661) (2.5, 0.9708) (3.0, 0.9748) (3.5, 0.9768) (4.0, 0.9783)};
                \addplot[icyan, mark=triangle*] coordinates {(0.5, 0.9183) (1.0, 0.9436) (1.5, 0.9601) (2.0, 0.9708) (2.5, 0.9753) (3.0, 0.9794) (3.5, 0.9815) (4.0, 0.9826)};
                \addplot[ired, mark=diamond*] coordinates {(0.5, 0.9252) (1.0, 0.9469) (1.5, 0.9632) (2.0, 0.9741) (2.5, 0.9786) (3.0, 0.9823) (3.5, 0.9843) (4.0, 0.9855)};
            \end{axis}
        \end{tikzpicture}
    \end{minipage}\hfill
    \begin{minipage}{0.33\textwidth}
        \centering
        \begin{tikzpicture}
            \begin{axis}[rowstyle, ylabel={UTMOS}, title={UTMOS (w/ SED w/ SSRR)},xlabel={Bitrate (kbps)}, xmin=0.25, xmax=4.25, ymin=2.0, ymax=3.4]
                \addplot[iblue, line width=1pt] coordinates {(0.0, 3.23) (4.5, 3.23)};
                \addplot[igreen, mark=square] coordinates {(0.5, 2.4045) (1.0, 2.9754) (1.5, 3.0031) (2.0, 3.0646) (2.5, 3.1198) (3.0, 3.1416) (3.5, 3.1744) (4.0, 3.1712)};
                \addplot[iorange, mark=o] coordinates {(0.5, 2.1980) (1.0, 2.9578) (1.5, 2.9920) (2.0, 3.0626) (2.5, 3.1115) (3.0, 3.1436) (3.5, 3.1582) (4.0, 3.1697)};
                \addplot[icyan, mark=triangle*] coordinates {(0.5, 2.2353) (1.0, 3.1581) (1.5, 3.1835) (2.0, 3.2550) (2.5, 3.2924) (3.0, 3.2898) (3.5, 3.3139) (4.0, 3.3229)};
                \addplot[ired, mark=diamond*] coordinates {(0.5, 2.3640) (1.0, 3.1837) (1.5, 3.2285) (2.0, 3.2620) (2.5, 3.2914) (3.0, 3.3084) (3.5, 3.3106) (4.0, 3.3167)};
            \end{axis}
        \end{tikzpicture}
    \end{minipage}
    \caption{Ablation studies on JHCodec performance. The top and middle rows analyze the impact of self-supervised representation reconstruction (SSRR) loss at 300k and 600k iterations, respectively, for RVQ with and without SED. The bottom row shows the effect of additional training for JHCodec with SSRR from 300k to 1.4M iterations.}
    \label{fig:full_ablation}
\end{figure*}
At 300k steps, before full convergence, the RVQ with SED achieves lower WER and CER than the RVQ without SED, both with and without SSRR cases, indicating greater robustness to incomplete optimization.
When SSRR is applied at 300k steps, both RVQ designs exhibit substantial gains in intelligibility and speaker similarity.
This improvement can be attributed to SSRR explicitly regularizing discrete representations to be invariant to self-reconstruction, thereby mitigating quantization noise and preventing unstable codebook assignments during early training.
As a result, SSRR reduces representation drift across frames and enforces more linguistically consistent tokenization, leading to more reliable downstream decoding.
Figure~\ref{fig:full_ablation}'s top row shows the effect of SSRR across different bitrates and RVQ configurations. 
Across all configurations, SSRR consistently improves intelligibility and speaker similarity.
At 300k steps, SSRR reduces WER by nearly half for both RVQ designs, highlighting its importance in stabilizing discrete representations under limited training budgets.

While SSRR may slightly reduce UTMOS in some settings, the overall trade-off is favorable, as the gains in intelligibility and speaker similarity outweigh minor decreases in perceptual quality, particularly because UTMOS scores remain above 3.17, while the ground truth's UTMOS is 3.23.



At 600k steps, the gap between models trained with and without SSRR remains substantial across both RVQ variants and all bitrates.
Without SSRR, both variants exhibit markedly higher WER and lower speaker similarity, whereas SSRR yields WERs close to the ground-truth reference at high bitrates.
This suggests that SSRR does not merely accelerate early convergence, but provides an explicit regularization effect that remains beneficial after longer optimization.
Furthermore, with SSRR, both RVQ designs achieve WERs within 1\% of the ground-truth WER after only 300k training steps, despite this being a very early stage of training.
Overall, these results demonstrate that SSRR improves the quality and stability of learned discrete representations and plays a more dominant role in reconstruction performance than previous losses, particularly in low- and mid-resource training regimes.

\begin{table*}[th]
    \caption{
    Comparisons with baseline codecs.
    $\dagger$ indicates that NanoCodec's encoder is non-streamable while the decoder is streamable.
     $\ddagger$ refers to the minimum theoretical latency of the model.
     $\ast$ denotes the latency measured after CUDA optimization, while all other latency results are measured in eager mode.
    JHCodec achieves the lowest training GPU budget, full streamability, the lowest latency, and a competitive real-time factor.
    }
    \label{tab:baseline}
    \centering
    \resizebox{0.99\linewidth}{!}{%
    \begin{tabular}{l|ccccccccccc}
    \toprule
     Model        
     & \# Parameters 
     & Training GPU Budget
     & Streamable
     & Lookahead
     & SED
     & SSRR
     & Bitrate (kbps) 
     & Frame Rate (Hz)
     & MAC (G)
     & Latency (ms)
     & RTF (Enc, Dec, Total) 
       \\
    \midrule
    Ground Truth (GT) &
    $-$ & $-$ & $-$ & $-$ & $-$ & $-$ & 256.\phantom{00} & $-$ & $-$ & $-$& $-$
     \\      
    \midrule
    \rowcolor{igray!20} \multicolumn{12}{l}{\textbf{\textit{Non-Streaming}}} \\
    \midrule
    DAC-8 & \phantom{0}75M & N/A & \xmark & $-$ & \xmark & \xmark & \phantom{00}4.00  & 50\phantom{.0} & 40.1 & $-$ & 0.0008, 0.0011, 0.0019 \\
    BigCodec & 159M & 8 A100 $\times$ 600k steps   & \xmark & $-$  & \xmark & \xmark & \phantom{00}1.04 & 80\phantom{.0} & 67.1 & $-$ & 0.0050, 0.0051, 0.0101 \\
    TAAE & 950M  & 16 H100 $\times$ 650k+ steps & \xmark & $-$  & \xmark & \cmark & \phantom{00}0.70 & 25\phantom{.0}  & 37.4& $-$ & 0.0019, 0.0020, 0.0039 \\
    NanoCodec & \phantom{0}62M &  48 A100  $\times$ 196k steps & \phantom{$^\dagger$}\xmark$^\dagger$ & $-$  & \xmark & \xmark & \phantom{00}1.78 & 12.5 & 48.5 & $-$ &0.0026, 0.0042, 0.0068  \\ 
    \midrule
    \rowcolor{igray!20} \multicolumn{12}{l}{\textbf{\textit{Streaming}}} \\
    \midrule
    Mimi-32 & \phantom{0}79M & 8 A100 $\times$ 1M steps & \cmark & \phantom{0}\textbf{0}ms  & \cmark & \xmark & \phantom{00}4.40 & 12.5  & \phantom{0}8.1 & \phantom{0}86.7 & 0.0012, 0.0008, 0.0020 \\
    FocalCodec-Stream & 249M & N/A & \cmark & 60ms  & \xmark & \xmark & \phantom{00}0.80 & 50\phantom{.0} & 13.5 & \phantom{$^\ddagger$0}80.0$^\ddagger$ & 0.0012, 0.0005, 0.0017 \\
    MagiCodec & 210M & N/A & \cmark&  20ms & \xmark & \xmark & \phantom{00}0.85 & 50\phantom{.0} & \phantom{0}7.1 & \phantom{$^\ddagger$0}60.0{$^\ddagger$} &  \textbf{0.0005}, \textbf{0.0004}, \textbf{0.0009}  \\
    \midrule
    \bf{JHCodec-8} & 271M & \begin{tabular}[t]{@{}c@{}}\textbf{1 H200 $\times$ 600k steps} \\ \textbf{+ 2 H200 $\times$ 800k steps}\end{tabular} &  \cmark & \phantom{0}\textbf{0}ms & \cmark & \cmark & \phantom{00}4.00 & 50\phantom{.0} & 13.6 & \textbf{26.8 (20.6)}$^\ast$  & \textbf{0.0006}, \textbf{0.0005}, \textbf{0.0011} \\
    \bottomrule
    \end{tabular}
    }
\end{table*}

The middle row of Figure~\ref{fig:full_ablation} compares models with and without SSRR after 600k steps.
The large gap between models with and without SSRR at the same 600k training budget across both RVQ styles demonstrates that SSRR does more than accelerate early convergence: it improves the absolute intelligibility and speaker-similarity performance attainable by the codec.
At 600k steps with SSRR, both RVQ variants achieve comparable WER and CER, while the Mimi-style RVQ consistently yields slightly higher S-SIM.
These results indicate that while the Mimi-style RVQ provides a stronger inductive bias in early training, the choice of RVQ becomes less critical as training progresses.
Each additional codebook increases the bitrate by 0.5 kbps.
While UTMOS already achieves sufficiently high scores (above 3), the model with Mimi-style RVQ consistently yields slightly better WER and S-SIM at all bitrates.
Based on these observations, we adopt the Mimi-style RVQ configuration for the final model.
The bottom row of Figure~\ref{fig:full_ablation} shows JHCodec with SSRR after 300k, 600k, 1M, and 1.4M training steps.
The 600k-step model, trained with a single GPU, already achieves competitive performance, while further training improves intelligibility and speaker similarity.



\subsection{Baselines}
We compare our method with both non-streaming and streaming neural audio codecs.
For non-streaming baselines, we evaluate DAC\footnote{\url{https://hf.co/descript/dac_16khz}} \cite{dac}, BigCodec\footnote{\url{https://hf.co/Alethia/BigCodec}} \cite{bigcodec}, TAAE\footnote{\url{https://hf.co/stabilityai/stable-codec-speech-16k}} \cite{stablecodec}, and NanoCodec\footnote{\url{https://hf.co/nvidia/nemo-nano-codec-22khz-1.78kbps-12.5fps}} \cite{nanocodec}.
For streaming evaluation, we include Mimi\footnote{\url{https://hf.co/kyutai/mimi}} \cite{moshi},
MagiCodec\footnote{\url{https://hf.co/Ereboas/MagiCodec_16k_50hz}} \cite{magicodec}, and FocalCodec-Stream\footnote{\url{https://hf.co/lucadellalib/focalcodec_50hz_65k_causal}} \cite{focalstream}.
While NanoCodec supports streaming at the decoder level, we categorize it as a non-streamable model since the overall system is not fully streamable.
Unlike recent work that compares only low-bitrate setups, we compared a wide variety of codecs, including high-bitrate RVQ codecs.
For RVQ-based codecs, the numeric suffix indicates the number of codebooks used.
For Mimi, both the commonly used Mimi-8 and the maximum-capacity Mimi-32 configurations are reported.
Table \ref{tab:baseline} shows the details of the baseline codecs.

Notably, prior codecs that report their training GPU budgets typically require more than 8 GPUs.
Our codec is trained with 1 H200 GPU for the first 600k steps and with 2 H200 GPUs for the remaining 800k steps. 
For brevity, we report the total training budget as the equivalent of 1 H200 GPU for 2.2M steps.

To evaluate computational efficiency, we measure the number of multiply-accumulate operations (MAC) for a 1-second audio input. 
For temporal metrics, including latency and real-time factor (RTF), measurements are conducted over a 10-second duration to ensure stability. 
All metrics are measured on an H200 GPU. 
The reported latency is comprehensive, encompassing the time required for input frame buffering, the lookahead window, and the actual model processing time. 
Moreover, Transformer-decoder-only models, MagiCodec, and our JHCodec, exhibit very low real-time factors (RTF), indicating fast speech resynthesis. 
While FocalCodec-Stream and MagiCodec did not offer an optimized streaming codebase, we mark their theoretical minimum latency. 
While other models show high latency due to long frame lengths and lookahead, JHCodec achieves the \textbf{lowest end-to-end latency} due to its high frame rate and zero lookahead.
As a result, JHCodec provides a practical advantage for real-time speech-to-speech systems, where minimizing codec latency is critical to overall latency.

As shown in Table~\ref{tab:opensource_streaming_codecs}, among the streaming neural audio codec baselines considered in this work, Mimi, FocalCodec-Stream, MagiCodec, and JHCodec differ substantially in their level of reproducibility.
While several models provide either inference code or pretrained checkpoints, JHCodec is the only open-source streaming codec baseline with the full training pipeline and streaming inference.
This makes JHCodec particularly suitable for reproducible research and further development, as it enables not only inference and evaluation but also full retraining and modification of the training objective.

\begin{table}[t]
\centering
\caption{Open-source availability of representative streaming neural audio codec baselines.}
\label{tab:opensource_streaming_codecs}
\centering
\resizebox{0.99\linewidth}{!}{%
\begin{tabular}{lcccc}
\toprule
Model & \multicolumn{2}{c}{Inference} & \multicolumn{2}{c}{Training} \\
\cmidrule(lr){2-3} \cmidrule(lr){4-5}
 & Checkpoint & Streaming & Dataset & Pipeline \\
\midrule
Mimi & \cmark & \cmark & \xmark & \xmark \\
FocalCodec-Stream & \cmark & \xmark & \cmark & \cmark \\
MagiCodec & \cmark & \xmark & \cmark & \xmark \\
\textbf{JHCodec} & \cmark & \cmark & \cmark &  \cmark \\
\bottomrule
\end{tabular}
}
\end{table}

\subsection{Downstream Automatic Speech Recognition}

To evaluate how well the discrete embedding preserves linguistic content, we trained automatic speech recognition (ASR) models using features extracted from the codec encoders and self-supervised models.
This test also evaluates the quality of SW2V's training.
We fine‑tune Whisper Small\footnote{\url{https://huggingface.co/openai/whisper-small}} \cite{whisper} on top of codec features and report WER. 
Inputs are projected into the Whisper input space with an adapter composed of two convolutional layers followed by two transformer adapter layers. We train only the adapter for one epoch, then unfreeze the last two encoder and decoder layers of Whisper. 
Training uses LibriSpeech train-clean splits (460h) \cite{librispeech} with a batch size of 32, a learning rate of $10^{‑4}$ with 500 warmup steps, and 20 epochs.
We tested models on the LibriSpeech test-clean dataset.

\section{Results and Discussion}
\subsection{LibriSpeech}
Table \ref{tab:results_main} reports intelligibility, speaker similarity, and perceptual quality metrics on the LibriSpeech test-clean subset.
Among non-streaming baselines, DAC-8 achieves strong intelligibility and speaker similarity, while BigCodec attains the highest perceptual quality in terms of UTMOS.
For non-streaming baselines, NanoCodec and DAC-8 show competitive WER and CER, while BigCodec achieves the highest UTMOS score.
Among the streaming baselines, Mimi-32 performs well on WER, CER, and S-SIM, ranking second-best in intelligibility metrics and best in speaker similarity, but yields lower UTMOS.
Our proposed JHCodec-8 achieves a superior balance across all metrics.
It ranks among the top-performing streaming models in WER, CER, and S-SIM while maintaining high perceptual quality. Notably, on LibriSpeech test-clean, JHCodec achieves the best WER and CER among all evaluated codecs, including non-streaming models, while also outperforming Mimi-32 and NanoCodec despite a significantly lower training budget.

While low-bitrate codecs generally exhibit slightly worse intelligibility metrics, BigCodec and MagiCodec achieve the highest perceptual quality among non-streaming and streaming, respectively.
One possible explanation is that low-bitrate representations limit the capacity to preserve detailed linguistic information.
As a result, models with limited complexity may prioritize perceptual fidelity over intelligibility.

\begin{table}[h]
    \caption{
    Evaluation results for JHCodec and baseline methods on the LibriSpeech test-clean dataset. 
    For both streaming and non-streaming models, the best and second-best results are indicated in \textbf{bold} and \underline{underline}, respectively.
    }
    \label{tab:results_main}
    \centering
    \resizebox{0.97\linewidth}{!}{%
    \begin{tabular}{l|cccc}
    \toprule
     Model        
     & WER ($\downarrow$)  
     & CER ($\downarrow$) 
     & S-SIM ($\uparrow$)  
     & UTMOS ($\uparrow$) 
       \\
    \midrule
    Ground Truth (GT) & 2.99 & 1.13 & 1.0000 & 3.2311 \\      
    \midrule
    \rowcolor{igray!20} \multicolumn{5}{l}{\textbf{\textit{Non-Streaming}}} \\
    \midrule
    DAC-8  & \underline{3.33}   &  \underline{1.29}   &  \underline{0.9832} & 2.5845 \\
    BigCodec&  3.67  &  1.50   & 0.9799 & \textbf{3.3694} \\
    TAAE  &  8.78   &  4.38   & 0.9371 & \underline{3.3495} \\
    NanoCodec   &  \textbf{3.16}   &   \textbf{1.24}   &  \textbf{0.9886} & 3.1630\\ 
    \midrule
    \rowcolor{igray!20} \multicolumn{5}{l}{\textbf{\textit{Streaming}}} \\
    \midrule
    Mimi-8 &  4.07   &  1.78   & 0.9673 & 2.7884  \\
    Mimi-32 & \underline{3.26}   &  \underline{1.29}   &  \textbf{0.9898} & 2.9685  \\
    FocalCodec-Stream &   4.05   &  1.66   & 0.9606 & 2.9772 \\
    MagiCodec & 4.35   &  1.85   & 0.9715 & \textbf{3.4901}  \\
    \bf{JHCodec-8}  &  \textbf{3.04}   &  \textbf{1.13} & \underline{0.9855} &  \underline{3.3167} \\
    \bottomrule
    \end{tabular}
    }
\end{table}

Table \ref{tab:results_test_other} reports results on the test-other subset. 
Although intelligibility degrades across all codecs, it is especially severe in low-bitrate codecs. 
While performance degrades across all models, the trends remain consistent. 
Among non-streaming baselines, the tendency is similar, but DAC achieves the best speaker similarity.
For streaming models, Mimi-32 achieves the best WER, CER, and speaker similarity, reflecting the benefit of higher RVQ capacity, while MagiCodec attains the highest UTMOS among streaming approaches.
Our proposed JHCodec-8 demonstrates competitive, well-balanced performance across all metrics, ranking among the second-best streaming models for intelligibility, speaker similarity, and perceptual quality.

\begin{table}[h]
    \caption{
    Evaluation results for the LibriSpeech test-other. 
    }
    \label{tab:results_test_other}
    \centering
    \resizebox{0.97\linewidth}{!}{%
    \begin{tabular}{l|cccc}
    \toprule
     Model        
     & WER ($\downarrow$)  
     & CER ($\downarrow$) 
     & S-SIM ($\uparrow$)  
     & UTMOS ($\uparrow$) 
       \\
    \midrule
    Ground Truth (GT) & \phantom{0}5.16 & 2.27 & 1.0000 & 2.9420 \\      
    \midrule
    \rowcolor{igray!20} \multicolumn{5}{l}{\textbf{\textit{Non-Streaming}}} \\
    \midrule
    DAC-8  &\phantom{0}\underline{6.23}   &  \underline{2.89}   & \textbf{0.9812} & 2.3838 \\
    BigCodec&  \phantom{0}8.22  &  4.10   & 0.9756 & \textbf{3.0748} \\
    TAAE  &  13.91 &  7.46   & 0.9312 & \underline{3.0550} \\
    NanoCodec   &  \textbf{\phantom{0}6.11}   &  \textbf{2.86}   & \underline{0.9682} & 2.8308 \\ 
    \midrule
    \rowcolor{igray!20} \multicolumn{5}{l}{\textbf{\textit{Streaming}}} \\
    \midrule
    Mimi-8 &  \phantom{0}9.62   &  4.95   & 0.9626 & 2.4902  \\
    Mimi-32 & \textbf{\phantom{0}5.83}   &  \textbf{2.66}   &  \textbf{0.9874} & 2.6489  \\
    FocalCodec-Stream &   \phantom{0}9.34  &  4.68   & 0.9536 & 2.7780 \\
    MagiCodec & 10.65   &  5.61   & 0.9665 & \textbf{3.2285}  \\
    \bf{JHCodec-8}  &  \phantom{0}\underline{6.01}   &  \underline{2.75} & \underline{0.9830} &  \underline{2.9998} \\
    \bottomrule
    \end{tabular}
    }
\end{table}

\subsection{TITW-Hard Test}
As shown in Table \ref{tab:results_titw}, all codecs exhibit substantial degradation for the TITW-Hard test set, particularly those at low bitrate codecs, where limited capacity struggles to resolve linguistic content from background noise.
Among non-streaming baselines, DAC-8 achieves the lowest dWER and dCER, indicating strong robustness in terms of intelligibility, while BigCodec and TAAE prioritize perceptual quality, as reflected by higher UTMOS scores.
For streaming models, Mimi-32 achieves the best intelligibility and speaker similarity, thanks to its higher RVQ capacity, whereas MagiCodec achieves the highest perceptual quality among streaming approaches. 
JHCodec-8 also demonstrates competitive and well-balanced performance across all metrics.
Compared with the LibriSpeech test-clean and test-other results, the same models remain the top two across all metrics, though their relative rankings change.

\begin{table}[h]
    \caption{
    Evaluation results for TITW-Hard test dataset. 
    }
    \label{tab:results_titw}
    \centering
    \resizebox{0.97\linewidth}{!}{%
    \begin{tabular}{l|cccc}
    \toprule
     Model        
     & dWER ($\downarrow$)  
     & dCER ($\downarrow$) 
     & S-SIM ($\uparrow$)  
     & UTMOS ($\uparrow$) 
       \\
    \midrule
    Ground Truth (GT) & \phantom{0}0.00 & \phantom{0}0.00 & 1.0000 & 2.5897 \\      
    \midrule
    \rowcolor{igray!20} \multicolumn{5}{l}{\textbf{\textit{Non-Streaming}}} \\
    \midrule
    DAC-8  & \textbf{11.70}   &  \textbf{\phantom{0}9.43}   & \underline{0.9628} & 2.1863 \\
    BigCodec&  18.77  &  14.58   & 0.9467 & \textbf{2.7134} \\
    TAAE  &  40.34   &  28.66   & 0.8224 & \underline{2.6022} \\
    NanoCodec   &  \underline{12.79}   &  \underline{10.25}   & \textbf{0.9683} & 2.5647 \\ 
    \midrule
    \rowcolor{igray!20} \multicolumn{5}{l}{\textbf{\textit{Streaming}}} \\
    \midrule
    Mimi-8 &  19.52   &  14.91   & 0.9296 & 2.3240  \\
    Mimi-32 & \textbf{11.57}  &  \textbf{\phantom{0}9.41}   & \textbf{0.9740} & 2.3998 \\
    FocalCodec-Stream & 20.23    & 15.42   & 0.9165 & 2.5293  \\
    MagiCodec &   20.10   &  15.37   & 0.9353 &  \textbf{2.9388} \\
    \bf{JHCodec-8}  &  \underline{11.98}   &  \phantom{0}\underline{9.49} & \underline{0.9635} &  \underline{2.6686} \\
    \bottomrule
    \end{tabular}
    }
\end{table}

\subsection{MLS-NonEnglish}
We evaluate cross-lingual generalization on the MLS-NonEnglish test splits. 
Since JHCodec and several baselines are trained exclusively on English data, this benchmark assesses whether the learned representations generalize to linguistic structures beyond the training distribution.
We report results for seven languages, denoted by their language codes: German (DE), Dutch (NL), French (FR), Italian (IT), Polish (PL), Portuguese (PT), and Spanish (ES).
Among non-streaming codecs, NanoCodec and DAC-8 demonstrate relatively strong generalization, while TAAE shows a significant degradation, indicating limited robustness to cross-lingual variability.
For streaming models, higher-capacity RVQ configurations such as Mimi-32 achieve the best intelligibility, suggesting that increased codebook capacity benefits multilingual reconstruction.
Notably, JHCodec-8 achieves competitive WER and CER among all baselines, ranking consistently among the top performers.
This result indicates that JHCodec’s discrete representations generalize reasonably well across languages, even though our model is trained exclusively on English speech.
Overall, these findings suggest that the proposed codec maintains robust linguistic preservation across languages while satisfying streaming constraints.

\begin{table}[h]
    \caption{%
        Language-specific WER (\%) results on the MLS non-English test splits.
        Overall denotes the aggregate WER across all evaluated utterances.
        Lower values are better.
    }
    \label{tab:results_mls_per_language}
    \centering
    \resizebox{0.98\linewidth}{!}{%
    \begin{tabular}{l|rrrrrrr|r}
        \toprule
        Model
        & DE
        & NL
        & FR
        & IT
        & PL
        & PT
        & ES
        & Overall \\
        \midrule
        Ground Truth (GT)
        & 5.51
        & 9.98
        & 5.21
        & 9.07
        & 4.46
        & 8.51
        & 4.04
        & 6.73 \\
        \midrule
        \rowcolor{igray!20}
        \multicolumn{9}{l}{\textbf{\textit{Non-Streaming}}} \\
        \midrule
        BigCodec
        & 8.69
        & 13.21
        & 8.11
        & 13.65
        & 8.24
        & 12.22
        & 5.88
        & 9.80 \\
        TAAE
        & 45.82
        & 55.11
        & 53.83
        & 53.94
        & 93.35
        & 71.35
        & 42.62
        & 52.72 \\
        DAC-8
        & \underline{6.41}
        & \underline{11.00}
        & \underline{5.98}
        & \underline{10.77}
        & \underline{5.56}
        & \underline{9.52}
        & \textbf{4.55}
        & \underline{7.64} \\
        NanoCodec
        & \textbf{6.28}
        & \textbf{10.45}
        & \textbf{5.88}
        & \textbf{10.99}
        & \textbf{5.32}
        & \textbf{9.15}
        & \underline{4.83}
        & \textbf{7.50} \\
        \midrule
        \rowcolor{igray!20}
        \multicolumn{9}{l}{\textbf{\textit{Streaming}}} \\
        \midrule
        Mimi-8
        & 9.94
        & 14.04
        & 8.79
        & 16.54
        & 11.07
        & 14.90
        & 8.52
        & 11.35 \\
        Mimi-32
        & \textbf{6.11}
        & \textbf{10.10}
        & \textbf{5.77}
        & \textbf{10.25}
        & \textbf{4.90}
        & \textbf{9.05}
        & \underline{5.00}
        & \textbf{7.30} \\
        FocalCodec-Stream
        & 9.96
        & 15.95
        & 9.14
        & 15.72
        & 10.77
        & 14.78
        & 6.59
        & 11.48 \\
        MagiCodec
        & 12.36
        & 18.25
        & 12.47
        & 18.72
        & 13.90
        & 16.42
        & 8.50
        & 13.96 \\
        \textbf{JHCodec-8}
        & \underline{6.26}
        & \underline{10.40}
        & \underline{5.89}
        & \underline{10.41}
        & \underline{5.26}
        & \underline{9.55}
        & \textbf{4.40}
        & \underline{7.38} \\
        \bottomrule
    \end{tabular}%
    }
\end{table}

\subsection{Downstream Automatic Speech Recognition}
Table \ref{tab:downstream_asr} summarizes the downstream ASR results. 
Among the non-streaming self-supervised representations, WavLM outperforms W2V-BERT 2.0, likely because we evaluate the full WavLM model, whereas W2V-BERT 2.0 uses a partial configuration, and because WavLM is trained solely on English data.
Similarly, our SW2V achieves the best WER among self-supervised representations, potentially reflecting training bias, yet indicating linguistic modeling capability.
Compared with other codecs, JHCodec demonstrates superior performance.


\begin{table}[h]
    \caption{
    Codec evaluation results for the downstream ASR. 
    }
    \label{tab:downstream_asr}
    \centering
    \resizebox{0.60\linewidth}{!}{%
    \begin{tabular}{l|cc}
    \toprule
     Model        
     & WER ($\downarrow$) & CER ($\downarrow$)  
       \\    
    \midrule
    Whisper Small &  \phantom{0}3.44  &   \phantom{0}1.24  \\  
    \midrule
    \rowcolor{igray!20} \multicolumn{3}{l}{\textbf{\textit{Non-Streaming}}} \\
    \midrule
    W2V2 (Full) & \phantom{0}\underline{4.73}  & \phantom{0}\textbf{2.17}   \\
    W2V-BERT 2.0 (17th) & \phantom{0}4.94 & \phantom{0}2.92  \\
    WavLM-Large (Full) & \phantom{0}\textbf{4.23} & \phantom{0}\underline{2.74}  \\
    \midrule
    DAC-8   & \phantom{0}\textbf{5.00} & \phantom{0}\textbf{2.60}   \\
    NanoCodec  &  \phantom{0}7.26  & \phantom{0}4.03  \\ 
    \midrule
    \rowcolor{igray!20} \multicolumn{3}{l}{\textbf{\textit{Streaming}}} \\
    \midrule
    \bf{SW2V} & \phantom{0}\textbf{4.11} & \phantom{0}\textbf{1.98}   \\
    \midrule
    Mimi-32 & \phantom{0}8.75 & \phantom{0}5.48  \\
    \bf{JHCodec-8}  & \phantom{0}\textbf{3.62} & \phantom{0}\textbf{1.50} \\
    \bottomrule
    \end{tabular}
    }
\end{table}

\subsection{General Remarks}

Several key observations emerge across the evaluations. 
First, regarding efficiency and low latency, JHCodec-8 achieves state-of-the-art performance, even outperforming Mimi-32 on clean data, despite a significantly lower training budget. 
Second, our analysis of the trade-off between intelligibility and quality reveals that while models such as BigCodec, TAAE, and MagiCodec often prioritize perceptual quality at the expense of WER, JHCodec provides a more consistent balance across both categories. 
Third, due to the ``semantic-acoustic conflict" observed in prior work, Mimi often achieves substantially lower UTMOS scores than GT across all test sets, whereas JHCodec, thanks to SSRR and denoising training, achieves UTMOS scores slightly higher than GT. 
Finally, DAC-8, NanoCodec, Mimi-32, and our JHCodec-8 achieve state-of-the-art performance, though their relative rankings vary across benchmarks.

\section{Conclusion}
Our results suggest that incorporating \textbf{self-supervised representation reconstruction} (SSRR) significantly benefits final performance and the training dynamics of streaming neural audio codecs. 
Far from being a simple auxiliary objective, SSRR appears to play a key role in balancing intelligibility and perceptual quality within a strict zero-lookahead streaming framework.
We show that SSRR substantially improves reconstruction intelligibility and accelerates convergence in our experiments. 
Competitive results are achieved within 300k steps on a single H200 GPU, and the final 1.4M-step model uses two H200 GPUs after 600k steps, while other models typically require larger-scale multi-node training setups to reach similar performance. 
This substantially lowers the computational barrier for future research in neural speech codecs, and we further support this goal by open-sourcing our full implementation.

Thanks to its zero-lookahead and low-latency architecture, the proposed JHCodec is particularly well-suited for real-time speech-to-speech systems.
Unlike alternative designs that rely on larger frame sizes or explicit lookahead, thereby increasing total system latency, we achieve competitive performance while strictly maintaining low end-to-end latency.

Furthermore, the proposed framework is not limited to speech codecs. 
The same principle can be extended to general audio codecs by leveraging universal audio representations trained on large-scale datasets \cite{samaudio}, potentially improving semantic consistency across broader acoustic domains.

\section*{Generative AI Use Disclosure}
We used generative AI to check grammar and improve the clarity of the submitted manuscript. 
We used generative AI for code auto-completion.
All generated text and code were reviewed by the authors before integration.
\section*{Acknowledgment}
This work was supported by the Office of the Director of National Intelligence (ODNI), Intelligence Advanced Research Projects Activity (IARPA), via the ARTS Program under contract D2023-2308110001. The views and conclusions contained herein are those of the authors and should not be interpreted as necessarily representing the official policies, either expressed or implied, of ODNI, IARPA, or the U.S. Government. The U.S. Government is authorized to reproduce and distribute reprints for governmental purposes notwithstanding any copyright annotation therein.

\bibliographystyle{IEEEtran}
\bibliography{mybib}

@article{moshi,
  title={Moshi: a speech-text foundation model for real-time dialogue},
  author={D{\'e}fossez, Alexandre and Mazar{\'e}, Laurent and Orsini, Manu and Royer, Am{\'e}lie and P{\'e}rez, Patrick and J{\'e}gou, Herv{\'e} and Grave, Edouard and Zeghidour, Neil},
  journal={arXiv preprint arXiv:2410.00037},
  year={2024}
}

@article{encodec,
  title={High fidelity neural audio compression},
  author={D{\'e}fossez, Alexandre and Copet, Jade and Synnaeve, Gabriel and Adi, Yossi},
  journal={arXiv preprint arXiv:2210.13438},
  year={2022}
}

@article{dac,
  title={High-fidelity audio compression with improved rvqgan},
  author={Kumar, Rithesh and Seetharaman, Prem and Luebs, Alejandro and Kumar, Ishaan and Kumar, Kundan},
  journal={NeurIPS},
  volume={36},
  pages={27980--27993},
  year={2023}
}

@article{vqgan,
  title={Vector-quantized image modeling with improved vqgan},
  author={Yu, Jiahui and Li, Xin and Koh, Jing Yu and Zhang, Han and Pang, Ruoming and Qin, James and Ku, Alexander and Xu, Yuanzhong and Baldridge, Jason and Wu, Yonghui},
  journal={arXiv preprint arXiv:2110.04627},
  year={2021}
}

@article{soundstream,
  title={Soundstream: An end-to-end neural audio codec},
  author={Zeghidour, Neil and Luebs, Alejandro and Omran, Ahmed and Skoglund, Jan and Tagliasacchi, Marco},
  journal={IEEE/ACM Transactions on Audio, Speech, and Language Processing},
  volume={30},
  pages={495--507},
  year={2021},
  publisher={IEEE}
}

@inproceedings{xcodec,
  title={Codec does matter: Exploring the semantic shortcoming of codec for audio language model},
  author={Ye, Zhen and Sun, Peiwen and Lei, Jiahe and Lin, Hongzhan and Tan, Xu and Dai, Zheqi and Kong, Qiuqiang and Chen, Jianyi and Pan, Jiahao and Liu, Qifeng and others},
  booktitle={Proceedings of the AAAI Conference on Artificial Intelligence},
  volume={39},
  number={24},
  pages={25697--25705},
  year={2025}
}

@article{xcodec2,
  title={Llasa: Scaling train-time and inference-time compute for llama-based speech synthesis},
  author={Ye, Zhen and Zhu, Xinfa and Chan, Chi-Min and Wang, Xinsheng and Tan, Xu and Lei, Jiahe and Peng, Yi and Liu, Haohe and Jin, Yizhu and Dai, Zheqi and others},
  journal={arXiv preprint arXiv:2502.04128},
  year={2025}
}

@article{xytokenizer,
  title={XY-Tokenizer: Mitigating the Semantic-Acoustic Conflict in Low-Bitrate Speech Codecs},
  author={Gong, Yitian and Jin, Luozhijie and Deng, Ruifan and Zhang, Dong and Zhang, Xin and Cheng, Qinyuan and Fei, Zhaoye and Li, Shimin and Qiu, Xipeng},
  journal={arXiv preprint arXiv:2506.23325},
  year={2025}
}

@article{valle,
  title={Neural codec language models are zero-shot text to speech synthesizers},
  author={Chen, Sanyuan and Wang, Chengyi and Wu, Yu and Zhang, Ziqiang and Zhou, Long and Liu, Shujie and Chen, Zhuo and Liu, Yanqing and Wang, Huaming and Li, Jinyu and others},
  journal={IEEE Transactions on Audio, Speech and Language Processing},
  year={2025},
  publisher={IEEE}
}

@article{speartts,
  title={Speak, read and prompt: High-fidelity text-to-speech with minimal supervision},
  author={Kharitonov, Eugene and Vincent, Damien and Borsos, Zal{\'a}n and Marinier, Rapha{\"e}l and Girgin, Sertan and Pietquin, Olivier and Sharifi, Matt and Tagliasacchi, Marco and Zeghidour, Neil},
  journal={Transactions of the Association for Computational Linguistics},
  volume={11},
  pages={1703--1718},
  year={2023},
  publisher={MIT Press One Broadway, 12th Floor, Cambridge, Massachusetts 02142, USA~…}
}

@article{w2v2,
  title={wav2vec 2.0: A framework for self-supervised learning of speech representations},
  author={Baevski, Alexei and Zhou, Yuhao and Mohamed, Abdelrahman and Auli, Michael},
  journal={NeurIPS},
  volume={33},
  pages={12449--12460},
  year={2020}
}

@article{w2vb2,
  title={SeamlessM4T: massively multilingual \& multimodal machine translation},
  author={Barrault, Lo{\"\i}c and Chung, Yu-An and Meglioli, Mariano Cora and Dale, David and Dong, Ning and Duquenne, Paul-Ambroise and Elsahar, Hady and Gong, Hongyu and Heffernan, Kevin and Hoffman, John and others},
  journal={arXiv preprint arXiv:2308.11596},
  year={2023}
}

@article{ts3codec,
  title={Ts3-codec: Transformer-based simple streaming single codec},
  author={Wu, Haibin and Kanda, Naoyuki and Eskimez, Sefik Emre and Li, Jinyu},
  journal={arXiv preprint arXiv:2411.18803},
  year={2024}
}

@article{magicodec,
  title={MagiCodec: Simple Masked Gaussian-Injected Codec for High-Fidelity Reconstruction and Generation},
  author={Song, Yakun and Chen, Jiawei and Zhuang, Xiaobin and Du, Chenpeng and Ma, Ziyang and Wu, Jian and Cong, Jian and Jia, Dongya and Chen, Zhuo and Wang, Yuping and others},
  journal={arXiv preprint arXiv:2506.00385},
  year={2025}
}

@article{wavlm,
  title={Wavlm: Large-scale self-supervised pre-training for full stack speech processing},
  author={Chen, Sanyuan and Wang, Chengyi and Chen, Zhengyang and Wu, Yu and Liu, Shujie and Chen, Zhuo and Li, Jinyu and Kanda, Naoyuki and Yoshioka, Takuya and Xiao, Xiong and others},
  journal={IEEE Journal of Selected Topics in Signal Processing},
  volume={16},
  number={6},
  pages={1505--1518},
  year={2022},
  publisher={IEEE}
}

@article{hubert,
  title={Hubert: Self-supervised speech representation learning by masked prediction of hidden units},
  author={Hsu, Wei-Ning and Bolte, Benjamin and Tsai, Yao-Hung Hubert and Lakhotia, Kushal and Salakhutdinov, Ruslan and Mohamed, Abdelrahman},
  journal={IEEE/ACM transactions on audio, speech, and language processing},
  volume={29},
  pages={3451--3460},
  year={2021},
  publisher={IEEE}
}

@article{speechtokenizer,
  title={Speechtokenizer: Unified speech tokenizer for speech large language models},
  author={Zhang, Xin and Zhang, Dong and Li, Shimin and Zhou, Yaqian and Qiu, Xipeng},
  journal={arXiv preprint arXiv:2308.16692},
  year={2023}
}

@inproceedings{phaseaug,
  title={{PhaseAug: a differentiable augmentation for speech synthesis to simulate one-to-many mapping}},
  author={Lee, Junhyeok and Han, Seungu and Cho, Hyunjae and Jung, Wonbin},
  booktitle={IEEE ICASSP},
  pages={1--5},
  year={2023},
  organization={IEEE}
}

@inproceedings{inconsistency,
  title={Analyzing and mitigating inconsistency in discrete speech tokens for neural codec language models},
  author={Liu, Wenrui and Guo, Zhifang and Xu, Jin and Lv, Yuanjun and Chu, Yunfei and Liu, Zemin and Lin, Junyang},
  booktitle={Proceedings of the 63rd Annual Meeting of the Association for Computational Linguistics (Volume 1: Long Papers)},
  pages={31035--31046},
  year={2025}
}

@article{hifigan,
  title={Hifi-gan: Generative adversarial networks for efficient and high fidelity speech synthesis},
  author={Kong, Jungil and Kim, Jaehyeon and Bae, Jaekyoung},
  journal={NeurIPS},
  volume={33},
  pages={17022--17033},
  year={2020}
}

@inproceedings{vits,
  title={Conditional variational autoencoder with adversarial learning for end-to-end text-to-speech},
  author={Kim, Jaehyeon and Kong, Jungil and Son, Juhee},
  booktitle={ICML},
  pages={5530--5540},
  year={2021},
  organization={PMLR}
}

@inproceedings{focal,
    title     = {{FocalCodec}: Low-Bitrate Speech Coding via Focal Modulation Networks},
    author    = {Luca {Della Libera} and Francesco Paissan and Cem Subakan and Mirco Ravanelli},
    booktitle = {NeurIPS},
    year      = {2025},
}

@article{focalstream,
  title={FocalCodec-Stream: Streaming Low-Bitrate Speech Coding via Causal Distillation},
  author={Della Libera, Luca and Subakan, Cem and Ravanelli, Mirco},
  journal={arXiv preprint arXiv:2509.16195},
  year={2025}
}

@article{wavtokenizer,
  title={Wavtokenizer: an efficient acoustic discrete codec tokenizer for audio language modeling},
  author={Ji, Shengpeng and Jiang, Ziyue and Wang, Wen and Chen, Yifu and Fang, Minghui and Zuo, Jialong and Yang, Qian and Cheng, Xize and Wang, Zehan and Li, Ruiqi and others},
  journal={arXiv preprint arXiv:2408.16532},
  year={2024}
}

@article{bigcodec,
  title={Bigcodec: Pushing the limits of low-bitrate neural speech codec},
  author={Xin, Detai and Tan, Xu and Takamichi, Shinnosuke and Saruwatari, Hiroshi},
  journal={arXiv preprint arXiv:2409.05377},
  year={2024}
}

@inproceedings{whisper,
  title={Robust speech recognition via large-scale weak supervision},
  author={Radford, Alec and Kim, Jong Wook and Xu, Tao and Brockman, Greg and McLeavey, Christine and Sutskever, Ilya},
  booktitle={ICML},
  pages={28492--28518},
  year={2023},
  organization={PMLR}
}

@article{musicgen,
  title={Simple and controllable music generation},
  author={Copet, Jade and Kreuk, Felix and Gat, Itai and Remez, Tal and Kant, David and Synnaeve, Gabriel and Adi, Yossi and D{\'e}fossez, Alexandre},
  journal={NeurIPS},
  volume={36},
  pages={47704--47720},
  year={2023}
}

@article{vqvae,
  title={Neural discrete representation learning},
  author={Van Den Oord, Aaron and Vinyals, Oriol and others},
  journal={NeurIPS},
  volume={30},
  year={2017}
}

@article{audiolm,
  title={Audiolm: a language modeling approach to audio generation},
  author={Borsos, Zal{\'a}n and Marinier, Rapha{\"e}l and Vincent, Damien and Kharitonov, Eugene and Pietquin, Olivier and Sharifi, Matt and Roblek, Dominik and Teboul, Olivier and Grangier, David and Tagliasacchi, Marco and others},
  journal={IEEE/ACM transactions on audio, speech, and language processing},
  volume={31},
  pages={2523--2533},
  year={2023},
  publisher={IEEE}
}

@inproceedings{gan,
  title={Generative Adversarial Nets},
  author={Ian J. Goodfellow and Jean Pouget-Abadie and Mehdi Mirza and Bing Xu and David Warde-Farley and Sherjil Ozair and Aaron C. Courville and Yoshua Bengio},
  booktitle={NeurIPS},
  year={2014},
}

@article{stablecodec,
  title={Scaling transformers for low-bitrate high-quality speech coding},
  author={Parker, Julian D and Smirnov, Anton and Pons, Jordi and Carr, CJ and Zukowski, Zack and Evans, Zach and Liu, Xubo},
  journal={arXiv preprint arXiv:2411.19842},
  year={2024}
}

@article{heptapod,
  title={Heptapod: Language Modeling on Visual Signals},
  author={Zhu, Yongxin and Chen, Jiawei and Chen, Yuanzhe and Chen, Zhuo and Jia, Dongya and Cong, Jian and Zhuang, Xiaobin and Wang, Yuping and Wang, Yuxuan},
  journal={arXiv preprint arXiv:2510.06673},
  year={2025}
}

@article{rvg1,
  title={When Worse is Better: Navigating the compression-generation tradeoff in visual tokenization},
  author={Ramanujan, Vivek and Tirumala, Kushal and Aghajanyan, Armen and Zettlemoyer, Luke and Farhadi, Ali},
  journal={arXiv preprint arXiv:2412.16326},
  year={2024}
}

@inproceedings{rvg2,
  title={Reconstruction vs. generation: Taming optimization dilemma in latent diffusion models},
  author={Yao, Jingfeng and Yang, Bin and Wang, Xinggang},
  booktitle={Proceedings of the Computer Vision and Pattern Recognition Conference},
  pages={15703--15712},
  year={2025}
}

@article{rvg3,
  title={Language Model Beats Diffusion--Tokenizer is Key to Visual Generation},
  author={Yu, Lijun and Lezama, Jos{\'e} and Gundavarapu, Nitesh B and Versari, Luca and Sohn, Kihyuk and Minnen, David and Cheng, Yong and Birodkar, Vighnesh and Gupta, Agrim and Gu, Xiuye and others},
  journal={arXiv preprint arXiv:2310.05737},
  year={2023}
}

@article{hibiki,
  title={High-fidelity simultaneous speech-to-speech translation},
  author={Labiausse, Tom and Mazar{\'e}, Laurent and Grave, Edouard and P{\'e}rez, Patrick and D{\'e}fossez, Alexandre and Zeghidour, Neil},
  journal={arXiv preprint arXiv:2502.03382},
  year={2025}
}

@inproceedings{flashattn,
 author = {Dao, Tri and Fu, Dan and Ermon, Stefano and Rudra, Atri and R\'{e}, Christopher},
 booktitle = {NeurIPS},
 pages = {16344--16359},
 title = {FlashAttention: Fast and Memory-Efficient Exact Attention with IO-Awareness},
 volume = {35},
 year = {2022}
}

@inproceedings{
flashattn2,
title={FlashAttention-2: Faster Attention with Better Parallelism and Work Partitioning},
author={Tri Dao},
booktitle={ICLR},
year={2024},
url={https://openreview.net/forum?id=mZn2Xyh9Ec}
}

@inproceedings{ss_phonetic,
  title     = {Self-Supervised Speech Representations are More Phonetic than Semantic},
  author    = {Kwanghee Choi and Ankita Pasad and Tomohiko Nakamura and Satoru Fukayama and Karen Livescu and Shinji Watanabe},
  year      = {2024},
  booktitle = {Interspeech},
}

@inproceedings{perceptual,
  title={Perceptual losses for real-time style transfer and super-resolution},
  author={Johnson, Justin and Alahi, Alexandre and Fei-Fei, Li},
  booktitle={European conference on computer vision},
  pages={694--711},
  year={2016},
  organization={Springer}
}

@inproceedings{preln,
  title={On layer normalization in the transformer architecture},
  author={Xiong, Ruibin and Yang, Yunchang and He, Di and Zheng, Kai and Zheng, Shuxin and Xing, Chen and Zhang, Huishuai and Lan, Yanyan and Wang, Liwei and Liu, Tieyan},
  booktitle={ICLR},
  year={2020},
}

@article{rotary,
  title={Roformer: Enhanced transformer with rotary position embedding},
  author={Su, Jianlin and Ahmed, Murtadha and Lu, Yu and Pan, Shengfeng and Bo, Wen and Liu, Yunfeng},
  journal={Neurocomputing},
  volume={568},
  year={2024},
  publisher={Elsevier}
}

@article{layernorm,
  title={Layer normalization},
  author={Ba, Jimmy Lei and Kiros, Jamie Ryan and Hinton, Geoffrey E},
  journal={arXiv:1607.06450},
  year={2016}
}

@article{rmsnorm,
  title={Root mean square layer normalization},
  author={Zhang, Biao and Sennrich, Rico},
  journal={NeurIPS},
  volume={32},
  year={2019}
}

@inproceedings{layerscale,
  title={Going deeper with image transformers},
  author={Touvron, Hugo and Cord, Matthieu and Sablayrolles, Alexandre and Synnaeve, Gabriel and J{\'e}gou, Herv{\'e}},
  booktitle={CVPR},
  year={2021}
}

@article{glu,
  title={Glu variants improve transformer},
  author={Shazeer, Noam},
  journal={arXiv preprint arXiv:2002.05202},
  year={2020}
}

@inproceedings{librittsr,
  title     = {LibriTTS-R: A Restored Multi-Speaker Text-to-Speech Corpus},
  author    = {Yuma Koizumi and Heiga Zen and Shigeki Karita and Yifan Ding and Kohei Yatabe and Nobuyuki Morioka and Michiel Bacchiani and Yu Zhang and Wei Han and Ankur Bapna},
  year      = {2023},
  booktitle = {Interspeech},
}

@article{ravdess,
  title={The Ryerson Audio-Visual Database of Emotional Speech and Song (RAVDESS): A dynamic, multimodal set of facial and vocal expressions in North American English},
  author={Livingstone, Steven R and Russo, Frank A},
  journal={PloS one},
  volume={13},
  number={5},
  year={2018},
}

@inproceedings{hifitts,
  author={Evelina Bakhturina and Vitaly Lavrukhin and Boris Ginsburg and Yang Zhang},
  title={{Hi-Fi Multi-Speaker English TTS Dataset}},
  year=2021,
  booktitle={Interspeech},
}

@INPROCEEDINGS{librilight,
  author={J. {Kahn} and M. {Rivière} and W. {Zheng} and E. {Kharitonov} and Q. {Xu} and P. E. {Mazaré} and J. {Karadayi} and V. {Liptchinsky} and R. {Collobert} and C. {Fuegen} and T. {Likhomanenko} and G. {Synnaeve} and A. {Joulin} and A. {Mohamed} and E. {Dupoux}},
  booktitle={IEEE ICASSP}, 
  title={Libri-Light: A Benchmark for ASR with Limited or No Supervision}, 
  year={2020},
}

@article{libriheavy,
      title={Libriheavy: a 50,000 hours ASR corpus with punctuation casing and context}, 
      author={Wei Kang and Xiaoyu Yang and Zengwei Yao and Fangjun Kuang and Yifan Yang and Liyong Guo and Long Lin and Daniel Povey},
    journal={arXiv:2309.08105},
    year={2023}
}

@inproceedings{mls,
  title     = {MLS: A Large-Scale Multilingual Dataset for Speech Research},
  author    = {Vineel Pratap and Qiantong Xu and Anuroop Sriram and Gabriel Synnaeve and Ronan Collobert},
  year      = {2020},
  booktitle = {Interspeech},
}

@misc{ljspeech,
  author       = {Keith Ito and Linda Johnson},
  title        = {The LJ Speech Dataset},
  howpublished = {\url{https://keithito.com/LJ-Speech-Dataset}},
  year         = 2017
}

@inproceedings{emilia,
  title={Emilia: An extensive, multilingual, and diverse speech dataset for large-scale speech generation},
  author={He, Haorui and Shang, Zengqiang and Wang, Chaoren and Li, Xuyuan and Gu, Yicheng and Hua, Hua and Liu, Liwei and Yang, Chen and Li, Jiaqi and Shi, Peiyang and others},
  booktitle={2024 IEEE SLT},
  pages={885--890},
  year={2024},
  organization={IEEE}
}

@misc{vctk,
  title={{Superseded-cstr vctk corpus: English multi-speaker corpus for cstr voice cloning toolkit(Version 0.92)}},
  author={Veaux, Christophe and Yamagishi, Junichi and MacDonald, Kirsten and others},
  year={2016},
  url={https://datashare.ed.ac.uk/handle/10283/3443},
  publisher={University of Edinburgh. The Centre for Speech Technology Research (CSTR)}
}

@inproceedings{perceptualse,
  title={Perceptual loss based speech denoising with an ensemble of audio pattern recognition and self-supervised models},
  author={Kataria, Saurabh and Villalba, Jes{\'u}s and Dehak, Najim},
  booktitle={IEEE ICASSP},
  pages={7118--7122},
  year={2021},
  organization={IEEE}
}

@article{effectivecontext,
  title={Effective Context in Neural Speech Models},
  author={Meng, Yen and Goldwater, Sharon and Tang, Hao},
  journal={arXiv preprint arXiv:2505.22487},
  year={2025}
}

@article{samaudio,
  title={SAM Audio: Segment Anything in Audio},
  author={Shi, Bowen and Tjandra, Andros and Hoffman, John and Wang, Helin and Wu, Yi-Chiao and Gao, Luya and Richter, Julius and Le, Matt and Vyas, Apoorv and Chen, Sanyuan and others},
  journal={arXiv preprint arXiv:2512.18099},
  year={2025}
}

@article{maskvct,
  title={MaskVCT: Masked Voice Codec Transformer for Zero-Shot Voice Conversion With Increased Controllability via Multiple Guidances},
  author={Lee, Junhyeok and Wang, Helin and Guan, Yaohan and Thebaud, Thomas and Moro-Velazquez, Laureano and Villalba, Jes{\'u}s and Dehak, Najim},
  journal={arXiv preprint arXiv:2509.17143},
  year={2025}
}

@article{pits,
  title={{PITS: Variational pitch inference without fundamental frequency for end-to-end pitch-controllable TTS}},
  author={Lee, Junhyeok and Jung, Wonbin and Cho, Hyunjae and Kim, Jaeyeon and Kim, Jaehwan},
  journal={arXiv preprint arXiv:2302.12391},
  year={2023}
}

@article{vevo,
  title={Vevo: Controllable zero-shot voice imitation with self-supervised disentanglement},
  author={Zhang, Xueyao and Zhang, Xiaohui and Peng, Kainan and Tang, Zhenyu and Manohar, Vimal and Liu, Yingru and Hwang, Jeff and Li, Dangna and Wang, Yuhao and Chan, Julian and others},
  journal={arXiv preprint arXiv:2502.07243},
  year={2025}
}

@article{melgan,
  title={Melgan: Generative adversarial networks for conditional waveform synthesis},
  author={Kumar, Kundan and Kumar, Rithesh and De Boissiere, Thibault and Gestin, Lucas and Teoh, Wei Zhen and Sotelo, Jose and De Brebisson, Alexandre and Bengio, Yoshua and Courville, Aaron C},
  journal={NeurIPS},
  volume={32},
  year={2019}
}

@inproceedings{codecsuperb,
  title={Codec-SUPERB: An in-depth analysis of sound codec models},
  author={Wu, Haibin and Chung, Ho-Lam and Lin, Yi-Cheng and Wu, Yuan-Kuei and Chen, Xuanjun and Pai, Yu-Chi and Wang, Hsiu-Hsuan and Chang, Kai-Wei and Liu, Alex and Lee, Hung-yi},
  booktitle={Findings of the Association for Computational Linguistics: ACL 2024},
  pages={10330--10348},
  year={2024}
}

@article{codecbench,
  title={CodecBench: A Comprehensive Benchmark for Acoustic and Semantic Evaluation},
  author={Deng, Ruifan and Gong, Yitian and Gao, Qinghui and Jin, Luozhijie and Cheng, Qinyuan and Fei, Zhaoye and Li, Shimin and Qiu, Xipeng},
  journal={arXiv preprint arXiv:2508.20660},
  year={2025}
}

@inproceedings{utmosv2,
  title     = {The T05 System for The {V}oice{MOS} {C}hallenge 2024: Transfer Learning from Deep Image Classifier to Naturalness {MOS} Prediction of High-Quality Synthetic Speech},
  author    = {Baba, Kaito and Nakata, Wataru and Saito, Yuki and Saruwatari, Hiroshi},
  booktitle = {IEEE SLT},
  year      = {2024},
}

@inproceedings{librispeech,
  title={Librispeech: an asr corpus based on public domain audio books},
  author={Panayotov, Vassil and Chen, Guoguo and Povey, Daniel and Khudanpur, Sanjeev},
  booktitle={IEEE ICASSP},
  pages={5206--5210},
  year={2015},
  organization={IEEE}
}

@article{titw,
  title={The Text-to-speech in the Wild (TITW) Database},
  author={Jung, Jee-weon and Zhang, Wangyou and Maiti, Soumi and Wu, Yihan and Wang, Xin and Kim, Ji-Hoon and Matsunaga, Yuta and Um, Seyun and Tian, Jinchuan and Shim, Hye-jin and others},
  journal={arXiv preprint arXiv:2409.08711},
  year={2024}
}

@inproceedings{
adamw,
title={Decoupled Weight Decay Regularization},
author={Ilya Loshchilov and Frank Hutter},
booktitle={ICLR},
year={2019},
}

@inproceedings{nanocodec,
  title     = {{NanoCodec: Towards High-Quality Ultra Fast Speech LLM Inference}},
  author    = {Edresson Casanova and Paarth Neekhara and Ryan Langman and Shehzeen Hussain and Subhankar Ghosh and Xuesong Yang and Ante Jukic and Jason Li and Boris Ginsburg},
  year      = {2025},
  booktitle = {{Interspeech}},
  pages     = {5028--5032},
  doi       = {10.21437/Interspeech.2025-827},
  issn      = {2958-1796},
}

@article{kvcache,
  title={Efficiently scaling transformer inference},
  author={Pope, Reiner and Douglas, Sholto and Chowdhery, Aakanksha and Devlin, Jacob and Bradbury, James and Heek, Jonathan and Xiao, Kefan and Agrawal, Shivani and Dean, Jeff},
  journal={Proceedings of machine learning and systems},
  volume={5},
  pages={606--624},
  year={2023}
}

@inproceedings{
flexicodec,
title={FlexiCodec: A Dynamic Neural Audio Codec for Low Frame Rates},
author={Jiaqi Li and Yao Qian and Yuxuan Hu and leying zhang and Xiaofei Wang and Heng Lu and Manthan Thakker and Jinyu Li and sheng zhao and Zhizheng Wu},
booktitle={ICLR},
year={2026},
url={https://openreview.net/forum?id=kYkfCs4ZAH}
}

@INPROCEEDINGS{stoi,
  author={Taal, Cees H. and Hendriks, Richard C. and Heusdens, Richard and Jensen, Jesper},
  booktitle={IEEE ICASSP}, 
  title={A short-time objective intelligibility measure for time-frequency weighted noisy speech}, 
  year={2010},
  volume={},
  number={},
  pages={4214-4217},
  doi={10.1109/ICASSP.2010.5495701}}

\end{document}